\documentclass[11pt]{article}

\usepackage[letterpaper,margin=1in]{geometry}
\usepackage[T1]{fontenc}
\usepackage[utf8]{inputenc}
\usepackage{lmodern}
\usepackage{microtype}
\usepackage{amsmath}
\usepackage{amssymb}
\usepackage{graphicx}
\usepackage{booktabs}
\usepackage{array}
\usepackage{tabularx}
\usepackage{enumitem}
\usepackage{xcolor}
\usepackage{caption}
\usepackage[section]{placeins}
\usepackage{hyperref}

\graphicspath{{images/}}
\hypersetup{
  colorlinks=true,
  linkcolor=blue!45!black,
  citecolor=blue!45!black,
  urlcolor=blue!45!black,
  pdftitle={When Is the Same Model Not the Same Service?},
  pdfauthor={Haorui Li et al.}
}

\setlist{nosep}
\newcommand{\aiping}{AI Ping}
\newcommand{\dataset}{the Q4 2025 \aiping{} measurement sample}
\newcommand{\figone}[2]{%
  \begin{figure}[!htbp]
    \centering
    \includegraphics[width=0.84\linewidth]{#2}
    \caption{#1}
    \label{fig:#2}
  \end{figure}
}
\newcommand{\figfour}[2]{%
  \begin{figure}[!htbp]
    \centering
    \includegraphics[width=0.455\linewidth]{#2-1}\hfill
    \includegraphics[width=0.455\linewidth]{#2-2}\\[0.35em]
    \includegraphics[width=0.455\linewidth]{#2-3}\hfill
    \includegraphics[width=0.455\linewidth]{#2-4}
    \caption{#1}
    \label{fig:#2}
  \end{figure}
}

\title{When Is the Same Model Not the Same Service?\\
A Measurement Study of Hosted Open-Weight LLM APIs}
\author{%
\begin{tabular}{c}
Haorui Li \quad Zhenghui He \quad Xuanzi Liu \quad Yang Xu \quad Dongsheng Liu\\
Jiakang Ma \quad Lupan Wu \quad Yangjie Wu \quad Xiongchao Tang \quad Tianhui Shi\textsuperscript{*}\\[0.45em]
\small QingCheng.AI AIPing Team\\
\small \textsuperscript{*}Corresponding author: Tianhui Shi (\texttt{sth@qingcheng.ai})
\end{tabular}
}
\date{May 2026}

\begin{document}
\maketitle

\begin{abstract}
Open-weight large language models (LLMs) are usually named as model artifacts, but production users often consume them as hosted API services. This paper argues that the operational unit is a \emph{service object}: a provider-specific, time-varying endpoint defined by model variant, protocol behavior, context capacity, listed price, latency and throughput distribution, reliability, and task feasibility. Using sampled request logs, provider metadata, compatibility probes, pricing snapshots, and continuous latency measurements collected by \aiping{} during Q4 2025, we study how this service layer changes the meaning of ``the same model.'' Three empirical patterns emerge. First, observed demand is concentrated but persistent across versions: in the displayed family aggregate, the largest family carries 32.0\% of relative demand and the top five carry 87.4\%, with a Gini coefficient of 0.693, while older variants remain active after newer releases. Second, supply and use separate: provider listing breadth does not imply realized adoption, and listed prices are more anchored than latency, throughput, context length, protocol support, and error semantics. Third, task mix matters: applications induce different token-length regimes, so provider choice is a constrained decision over provider-model-task-time tuples rather than a lookup by model name. In two representative counterfactuals under observed feasibility constraints, routing lowers Qwen3-32B cost by 37.8\% and raises DeepSeek-V3.2 average throughput by about 90\% relative to direct official access. The results support a measurement view of hosted open-weight LLMs as heterogeneous services, not static catalog entries. We open-source the measurement methodology and reproduction artifacts at \url{https://github.com/haoruilee/llm_api_measurement_study} to support result reproduction.
\end{abstract}

\section{Introduction}
Modern LLM deployment no longer ends at a model release. Open-weight models such as DeepSeek, Qwen, Kimi, GLM, and MiniMax are not only artifacts for local inference; they also become shared endpoints served by multiple providers with different prices, capacity plans, protocol implementations, context windows, safety wrappers, quantization choices, and latency profiles. The same nominal model can therefore expose different operational behavior depending on where and when it is called.

This paper focuses on the cross-provider API layer that sits between model capability and application behavior. We define a hosted open-weight LLM \emph{service object} as
\[
  S=(v,p,\pi,W,C,L,R,t),
\]
where \(v\) is the model variant or checkpoint label, \(p\) is the provider, \(\pi\) is the protocol surface, \(W\) is the effective context policy, \(C\) is the price schedule, \(L\) is the latency and throughput distribution, \(R\) is reliability and error recoverability, and \(t\) is measurement time. This definition separates a model artifact from the service actually consumed by applications.

The measurement target is not whether one model is intrinsically better than another, but how nominally identical or closely related open-weight models behave once they are exposed through independently operated API endpoints. We distinguish three identity layers throughout the paper: checkpoint-level claims require evidence about the exact released weights; variant-level claims concern named provider variants such as instruction, thinking, or dated releases; family-level claims aggregate related variants only for distributional summaries. This distinction is central because many public dashboards and API catalogs collapse these layers, while production behavior often depends on the finer service object.

Using \dataset{}, we study request volume, provider coverage, listed prices, time to first token (TTFT), tokens per second (TPS), context length, protocol compatibility, slow-response behavior, task category, and routing policy. The dataset is a sampled measurement view from \aiping{} rather than a census of all LLM API activity; the results should therefore be read as evidence about the observed service layer, not as global market shares.

The main finding is that model name is not a sufficient operational unit. Demand is concentrated, but older variants remain in use; provider coverage is broad, but realized usage is uneven; prices are often anchored, but latency, throughput, protocol behavior, context capacity, and error recoverability vary enough to change application outcomes. Routing has measurable value in the sample because it exploits cross-provider dispersion under feasibility constraints, not because any provider is universally superior. Throughout the paper, directly observed aggregate facts are separated from mechanism-level interpretations suggested by the data.

\section{Background and Related Work}
Recent model technical reports document the rapid improvement of open-weight LLM capability. The open-weight families appearing in or adjacent to this study are documented by DeepSeek-V2~\cite{deepseekv2}, DeepSeek-V3~\cite{deepseekv3}, DeepSeek-R1~\cite{deepseekr1}, Qwen~\cite{qwen}, Qwen2.5~\cite{qwen25}, Qwen3~\cite{qwen3}, Kimi K2~\cite{kimik2}, MiniMax-01~\cite{minimax01}, MiniMax-M1~\cite{minimaxm1}, and GLM-4.5~\cite{glm45}. These reports primarily describe model architectures, training data, post-training procedures, benchmark performance, context capability, or reasoning behavior. Their natural unit of analysis is the model family or checkpoint. This paper instead asks what happens after those models enter a multi-provider API market, where the same nominal model can be exposed through different infrastructure and protocol layers.

The systems literature studies how LLM serving systems manage batching, scheduling, memory, and throughput. Orca introduced iteration-level scheduling and selective batching for transformer serving~\cite{orca}; PagedAttention improves KV-cache memory management for high-throughput serving~\cite{pagedattention}; SGLang optimizes structured generation workloads through a co-designed frontend and runtime~\cite{sglang}; and DistServe separates prefill and decode resources to optimize latency-constrained serving~\cite{distserve}. Recent surveys summarize the broader design space of LLM inference serving~\cite{llmservingsurvey}. This paper is complementary: it does not propose a serving system, but measures externally visible consequences of heterogeneous serving deployments.

A related line of work studies cost-quality trade-offs and model routing. FrugalGPT formulates cascades for reducing LLM inference cost while preserving quality~\cite{frugalgpt}, and RouteLLM learns to route requests between models using preference data~\cite{routellm}. Our setting differs in that routing occurs not only across models but also across providers for the same model family. The decision boundary therefore includes context-window feasibility, protocol compatibility, latency tails, throughput, error semantics, and time-varying provider behavior.

Two other lines of work are especially relevant to our measurement boundary. First, Model Equality Testing formulates the question ``which model is this API serving?'' as a two-sample testing problem, motivated by the possibility that providers may quantize, watermark, fine-tune, or otherwise modify an advertised model~\cite{modelequality}. We do not claim checkpoint-level equality for every provider endpoint in this paper. Instead, we treat model identity as a measured boundary condition and report family-level or variant-level summaries only where the evidence supports them. Second, large-scale usage studies such as the OpenRouter 100T token study and BurstGPT emphasize that real production traffic differs from benchmark workloads and that request metadata, timing, model/provider identifiers, token counts, and cancellation behavior define the observable sample frame~\cite{openrouterstateai,burstgpt}. Our methodology follows this measurement style but focuses specifically on hosted open-weight API service heterogeneity.

\section{Measurement Methodology}
\subsection{Dataset and Units of Analysis}
The study combines several observation units over the Q4 2025 measurement window (2025-10-01 to 2025-12-31, Beijing time). At the model level, we distinguish checkpoint labels, provider-visible variants, and family aggregates. Family aggregation is used only for public distributional summaries and follows the rule in Appendix~\ref{app:classification}. At the provider level, we observe whether a provider lists a model, its relative price, its advertised or observed context length, and its measured service behavior. At the request level, we observe calls routed through the measurement environment, including token length, task category when available, selected routing policy, and timing statistics.

The resulting design is observational rather than randomized. Provider choice, model choice, and task mix are influenced by application needs. We therefore avoid causal claims such as ``provider \(A\) causes lower latency than provider \(B\)'' unless a specific counterfactual comparison is defined on the same model and similar request population. Most findings are descriptive distributional claims.

Table~\ref{tab:sample_frame} summarizes the public sample frame. Some cells intentionally report scope rather than raw counts because the underlying request logs include customer-sensitive operational traces. This boundary is weaker than a fully open measurement trace, but it is preferable to reporting unlabeled dashboard values without a data model.

\begin{table}[t]
\centering
\caption{Public sample-frame description for \dataset{}. The table defines what each observation layer means in the paper.}
\label{tab:sample_frame}
\footnotesize
\setlength{\tabcolsep}{4pt}
\renewcommand{\arraystretch}{1.12}
\begin{tabularx}{\linewidth}{l X}
\toprule
\textbf{Layer} & \textbf{Public definition} \\
\midrule
Time window & 2025-10-01 to 2025-12-31, Beijing time. \\
Provider universe & Providers visible to the \aiping{} service layer during the measurement window; provider-count figures use the observed listing set, while compatibility probes use the tested subset. \\
Model identity & Checkpoint and provider-visible variant names are preserved when available; family aggregates are used for public summary figures and are marked as such. \\
Request events & API calls visible to the measurement environment, with model/provider identifiers, timing fields, token counts, completion status when available, routing policy, and task label when available. \\
Demand denominator & Unless otherwise stated, demand figures refer to request-count mass in the displayed public aggregate, normalized by a common constant. \\
Reliability denominator & Slow-response and error-semantics figures use provider-model observations for which timing or error fields are available. \\
Task categories & Task labels are used only for the labeled subset and are not assumed to represent unlabeled traffic. \\
\bottomrule
\end{tabularx}
\end{table}

\subsection{Metric Definitions and Normalization}
Request-volume figures are normalized where required to avoid disclosing absolute traffic. Provider counts are reported as observed counts in the sampled provider set. Price comparisons use listed prices collected from provider websites at historical points; these prices may differ from private contracts. Performance is summarized using TTFT and TPS. Context length is treated as a capacity variable because falling below an application's required context length can make an endpoint infeasible even if its price and speed are acceptable.

The compatibility study tests whether providers implement selected OpenAI-compatible and Anthropic-compatible interfaces for representative models. Interface quality is analyzed separately from raw availability because a nominally supported API can still fail through incorrect status codes, empty error bodies, streaming error payloads, unsupported parameters, or unstable field schemas.

For concentration statistics, the public paper reports point estimates computed over the displayed normalized model-family set. These estimates are not population parameters for all LLM traffic. For a full reproducibility package, each headline statistic should be accompanied by the denominator, aggregation level, and a nonparametric bootstrap interval over the relevant unit: requests for demand shares, provider-model pairs for compatibility and context coverage, and measurement runs or request buckets for latency and throughput. When those denominators are not public, this manuscript treats the number as a descriptive statistic rather than a precise market estimate.

\subsection{Latency, Throughput, and Compatibility Measurements}
For streaming calls, TTFT is measured as the elapsed wall-clock time between request submission and receipt of the first output token. TPS is measured during the generation phase as output tokens divided by generation time after the first token. End-to-end latency includes both prefill and decode phases and is therefore affected by input length, output length, queuing, provider-side batching, and network delay. Context length is taken from provider-published limits or from endpoint behavior when the published limit and observed behavior differ. Compatibility tests use protocol-specific request templates and record whether the provider accepts the protocol, returns the expected schema, and exposes stable error semantics under failure.

Slow-response rates are computed at the provider-model granularity against an operational latency threshold used by the measurement system. Because the threshold is not yet part of the public artifact, these figures should be read as operational categories rather than independently reproducible latency estimates. A stronger release should publish either the threshold or threshold-free summaries such as p50, p90, p95, and p99 end-to-end latency, TTFT, and TPS by provider-model pair. This boundary is important for interpreting the results: the paper is intended to make the observed statistical structure visible without exposing customer-specific usage volumes or operational traces.

\subsection{Research Questions}
We organize the analysis around four statistical questions:
\begin{enumerate}
  \item How concentrated is model demand, and how persistent are older model versions after newer releases?
  \item Does provider-side model coverage match request-side demand?
  \item Along which dimensions do providers differ most: price, latency, throughput, context length, compatibility, or error behavior?
  \item How do task-level token distributions condition the optimal trade-off between cost, latency, throughput, and context length?
\end{enumerate}

\subsection{A Decision-Theoretic View of Provider Selection}
For a request \(r\) using model \(m\), let \(p\) index providers and let \(a(r)\) denote the task category when available. We treat protocol support and context length as feasibility constraints:
\[
  \mathcal{F}(r,m)=\{p: \operatorname{protocol}_{p,m}(r)=1,\; W_{p,m}\geq w_r,\; \operatorname{available}_{p,m}=1\},
\]
where \(W_{p,m}\) is the supported context window and \(w_r\) is the request's required context. Conditional on this feasible set, provider selection can be written as a request-dependent loss minimization problem:
\[
  p^\star(r)=\arg\min_{p\in\mathcal{F}(r,m)}
  \mathbb{E}\!\left[
    \alpha_{a(r)}\,\operatorname{TTFT}_{p,m,t}
    +\beta_{a(r)}\,\operatorname{Cost}_{p,m}(r)
    -\gamma_{a(r)}\,\operatorname{TPS}_{p,m,t}
    +\delta_{a(r)}\,\operatorname{Risk}_{p,m,t}
  \right].
\]
The task-dependent weights \(\alpha,\beta,\gamma,\delta\) encode the empirical fact that a short interactive request, a long-output generation request, and a long-context retrieval request do not optimize the same metric. The risk term is not a residual nuisance term; it decomposes into timeout risk, malformed-schema risk, context-truncation risk, billing risk, and model-identity or quality-drift risk. In deployments where an application requires strict output format or tool-call compatibility, those constraints should be moved from the loss function into the feasible set.

Equivalently, routing can be written as constrained optimization:
\[
\begin{aligned}
  \min_{p\in\mathcal{F}(r,m)} \quad &
    \mathbb{E}\left[\operatorname{Cost}_{p,m}(r)\right]
    \quad\text{or}\quad
    \mathbb{E}\left[\operatorname{Latency}_{p,m,t}(r)\right] \\
  \text{s.t.}\quad &
    \Pr(\operatorname{success}_{p,m,t})\geq \eta,\\
  & \operatorname{quality}_{p,m,a(r)}\geq q,\\
  & W_{p,m}\geq w_r,\quad
    \operatorname{protocol}_{p,m}(r)=1,\quad
    \operatorname{recoverable}_{p,m}=1 .
\end{aligned}
\]
The present public data support cost, speed, context, protocol, and error-semantics analysis more directly than semantic quality equivalence. The routing results below should therefore be interpreted as cost/throughput counterfactuals under observed feasibility constraints, not as proof that all routed providers return quality-equivalent completions.

This framing also explains why aggregate provider rankings can be misleading. The relevant random variable is not a provider-level scalar but a provider-model-task-time tuple. Aggregating across task categories can hide conditional specialization, and aggregating across time can hide service-quality drift. Routing gains arise when the conditional distribution across feasible providers has nonzero dispersion: selecting a provider from the lower tail of latency or cost, or the upper tail of throughput, produces an order-statistic gain relative to a fixed baseline. The counterfactual comparisons reported below estimate this average gain for two representative model-specific samples.

\section{Model Demand: Concentration, Persistence, and Coverage}
\subsection{Concentration and Version Persistence}
Figure~\ref{fig:fig1_model_demand} shows normalized request volume for frequently used open-weight models. The distribution is head-heavy: DeepSeek-V3/R1 leads the sample, followed by DeepSeek-V3.2, while high-volume Qwen variants such as Qwen3-32B, Qwen2.5-72B, and Qwen3-235B-A22B occupy the next tier. This pattern is consistent with a rank-frequency distribution in which a small number of model families account for a large fraction of observed API calls.

\figone{Model-demand concentration in the displayed aggregate. Panel A shows request volume normalized to DeepSeek-V3/R1. Panel B shows the Lorenz curve over the displayed model-family set; the shaded area indicates deviation from a uniform distribution.}{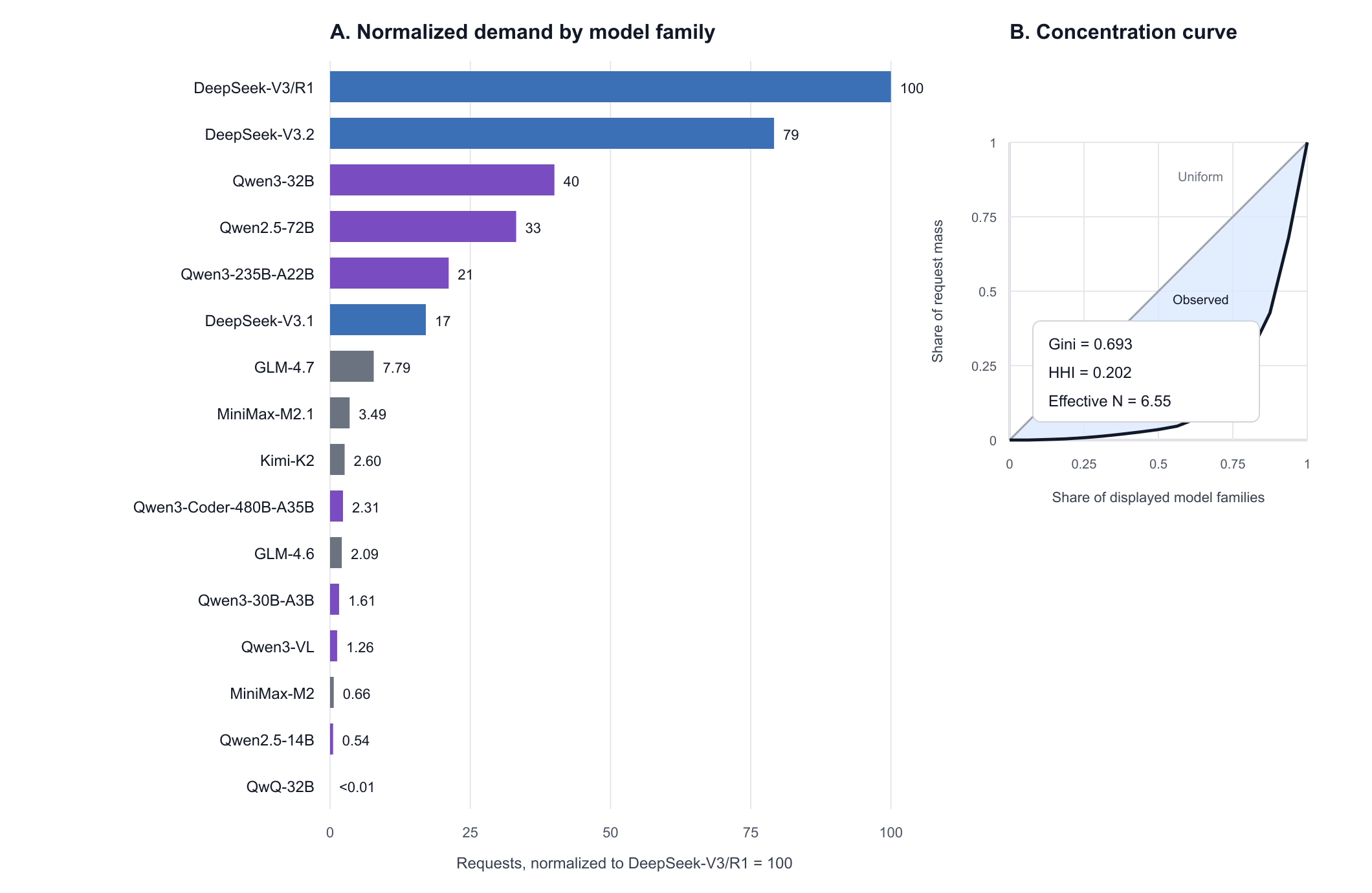}

To quantify this concentration, we compute scale-invariant inequality indices from the normalized bars displayed in Figure~\ref{fig:fig1_model_demand}. Let \(x_i\) denote the displayed normalized request volume for model family \(i\), and \(s_i=x_i/\sum_j x_j\) be the relative mass among the displayed families. Because all \(x_i\) are normalized by the same constant, these concentration indices are unchanged by the public normalization. Table~\ref{tab:concentration} reports the resulting values.

\begin{table}[t]
\centering
\caption{Concentration statistics computed from the normalized model-demand values shown in Figure~\ref{fig:fig1_model_demand}. Values describe the displayed model-family set, not the entire private request population.}
\label{tab:concentration}
\begin{tabular}{l r}
\toprule
\textbf{Statistic} & \textbf{Value} \\
\midrule
Top-1 relative mass & 0.320 \\
Top-2 relative mass & 0.573 \\
Top-3 relative mass & 0.700 \\
Top-5 relative mass & 0.874 \\
Herfindahl--Hirschman index & 0.202 \\
Gini coefficient & 0.693 \\
Entropy effective number, \(\exp(-\sum_i s_i\log s_i)\) & 6.55 \\
\bottomrule
\end{tabular}
\end{table}

The concentration statistics show that the displayed demand is not merely ranked but sharply unequal. The top five model families account for 87.4\% of displayed relative mass, while the entropy effective number is only 6.55 despite 16 displayed families. The more informative result, however, is not simply which model ranks first, but the coexistence of versions. Qwen2.5-72B remains visible despite later Qwen releases, and QwQ-32B is not immediately displaced by Qwen3-32B. Similarly, newer DeepSeek-V3.1 and DeepSeek-V3.2 variants do not fully remove DeepSeek-V3/R1 traffic. This is evidence of version persistence in the observed sample. It is consistent with production adoption inertia, but the stronger inertia claim would require release-aligned time series, cohort migration curves, or survival analysis of version replacement. A plausible mechanism is that a model version that has passed application-specific validation, prompt tuning, regression tests, and cost controls can remain a stable baseline after nominally stronger versions appear. In statistical terms, previous integration state is a likely covariate of current model choice, but the present cross-sectional figure alone should not be overread as a hazard model of model replacement.

\subsection{Supply Coverage Across Model Families}
Figure~\ref{fig:2} reports the number of service providers that offer API access to each model family. DeepSeek models have the broadest observed support. In the sampled provider universe, \texttt{aiping.cn} lists 29 providers; DeepSeek-V3/R1 and DeepSeek-V3.1 are supported by 23 providers each, and 24 providers support at least one DeepSeek model. The difference between support for individual variants and support for a family shows that provider coverage should be analyzed at both version and family levels.

\figone{Number of observed providers offering API access to each model family.}{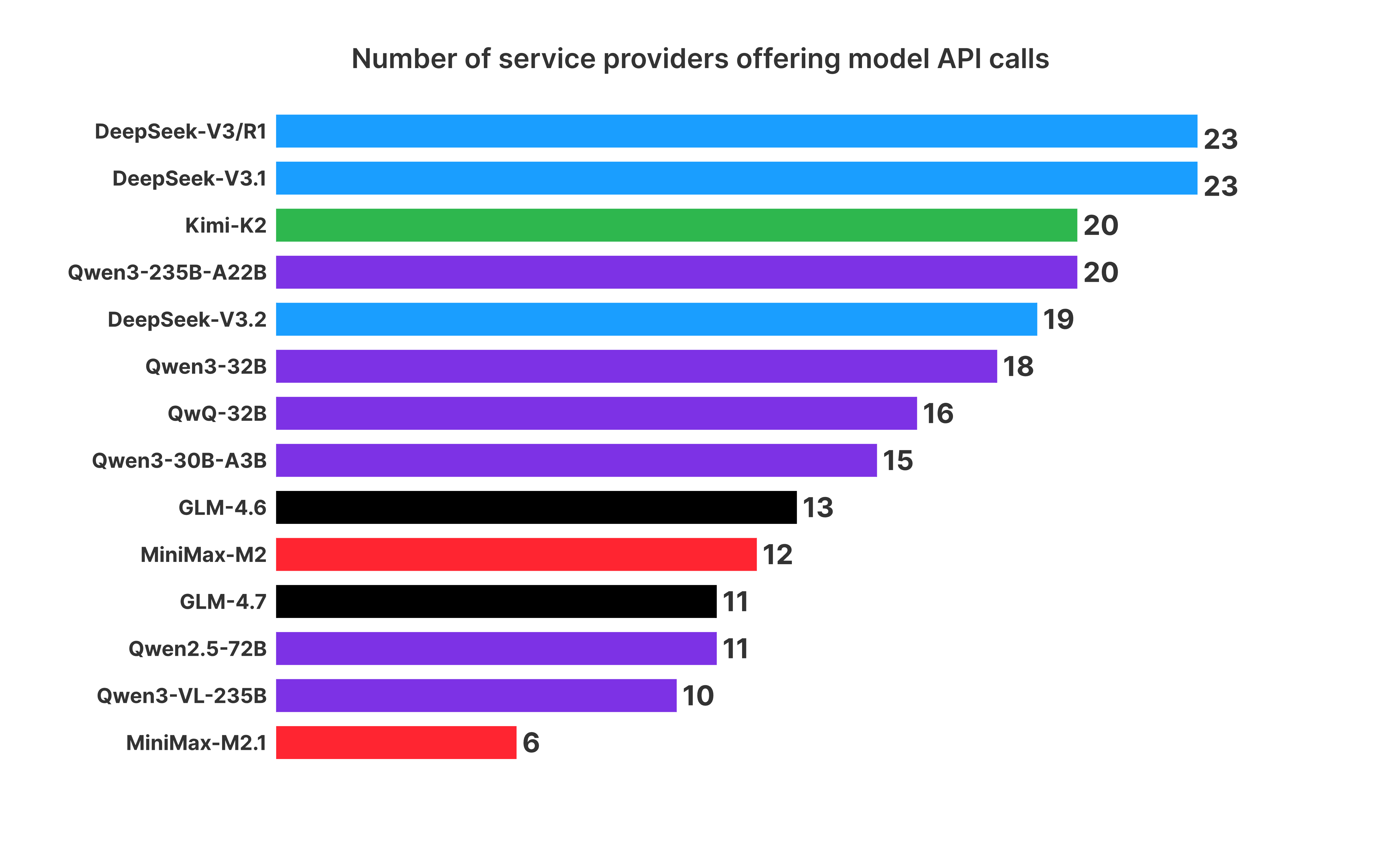}

Coverage is also stratified within model families. Figure~\ref{fig:3-1} shows that DeepSeek and Qwen variants have different provider availability. DeepSeek-V3.1 and DeepSeek-R1-0528 have higher coverage than some older or branch variants. For Qwen, base variants generally have broader coverage than instruction-tuned or thinking variants. This creates a supply-side gradient: providers first support widely deployable general variants, while demand-side applications often require instruction-following or reasoning variants for specific tasks. The resulting mismatch is not a binary shortage, but a conditional shortage: a model family may be widely available while the exact variant needed by an application remains less available.

\figone{Provider coverage for DeepSeek and Qwen variants.}{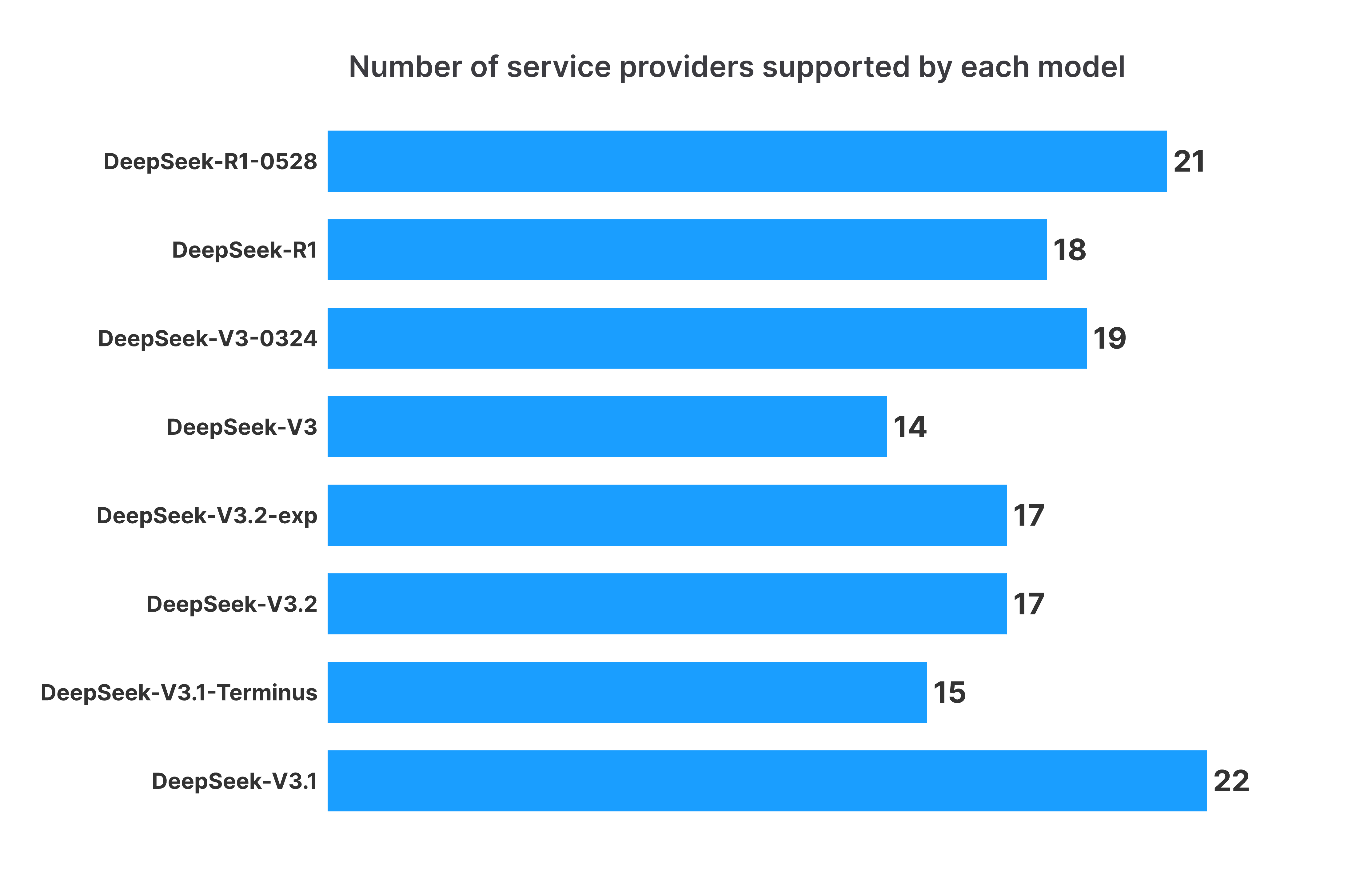}

\subsection{Coverage Does Not Imply Adoption}
Figure~\ref{fig:4} compares model call volume with the number of providers offering the model. The head of the demand distribution tends to coincide with broad provider support, but the association is not linear. For example, QwQ-32B reaches a nontrivial provider count while receiving limited observed traffic. This implies that provider coverage is a necessary but insufficient condition for demand. Demand also depends on release timing, task suitability, migration cost, price expectations, ecosystem familiarity, and the amount of application validation already invested in a model.

\figone{Model request volume, normalized against DeepSeek-V3, versus provider count.}{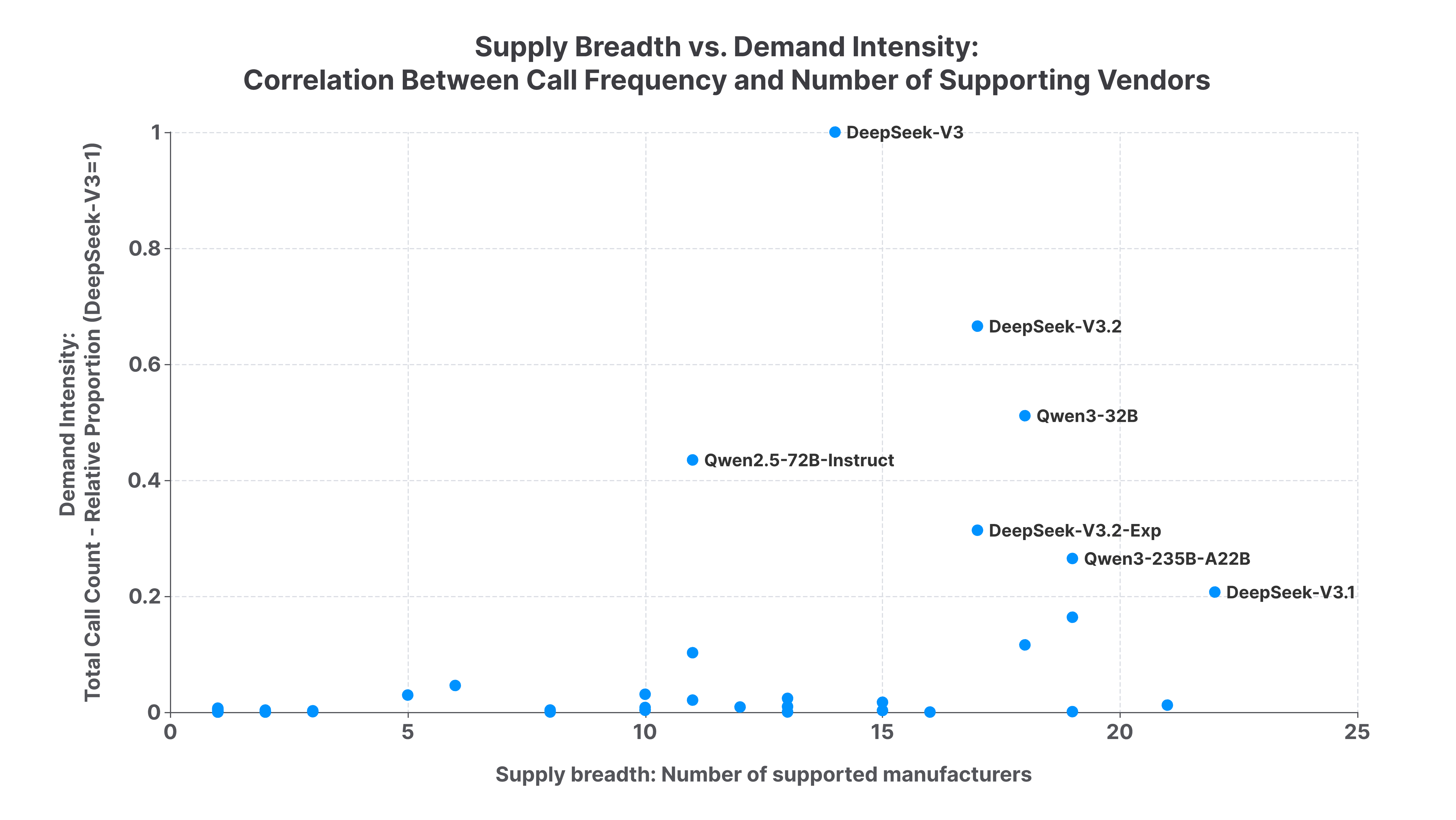}

This observation matters for interpreting ecosystem maturity. A model with many providers but low usage indicates supply anticipation or stranded availability. A model with high usage and few providers indicates concentration risk. A model with both high usage and high provider count forms the statistically stable core of the observed ecosystem.

\subsection{Parameter Scale Is Not a Sufficient Price Model}
Figure~\ref{fig:5} compares input/output prices with total and active parameter counts for selected large models. The sample does not support a simple monotone interpretation of price as a function of total parameter count. In mixture-of-experts (MoE) architectures, total parameter count measures stored capacity, while active parameters better approximate per-token computation. Even active parameter count, however, does not fully determine price: similar active scales can exhibit different listed prices across model families and providers.

\figfour{Parameter scale and listed API prices for selected large models.}{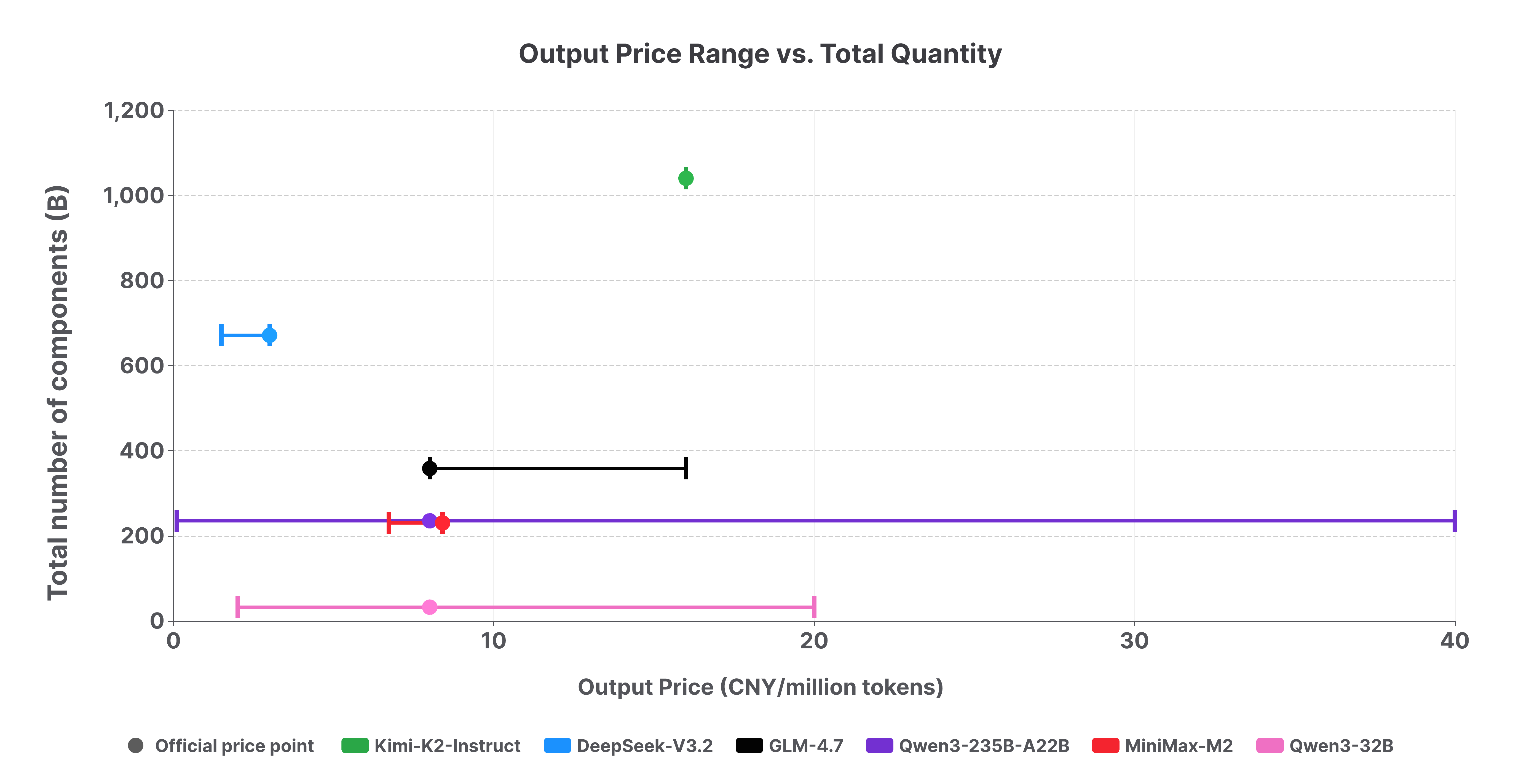}

The price distribution should therefore be interpreted as a compound outcome of architecture, provider operating cost, capacity planning, official reference prices, provider positioning, and possible volume discounts. Listed prices cluster near official benchmarks for many models, which suggests price anchoring. The more statistically important dispersion appears later in the performance metrics: when prices are compressed but latency and throughput vary widely, provider choice becomes a quality-variance problem rather than a pure price-minimization problem.

\section{Provider Heterogeneity}
\subsection{Breadth of Model Support}
Figure~\ref{fig:6-1} compares model richness across providers. The leading providers support a similar number of open-weight models, but they are not equivalent: two providers with the same model count can support different variants, context windows, protocols, and performance levels. Model count is therefore a weak scalar summary. It measures breadth, but not depth, reliability, or compatibility.

\figone{Number of open-weight models supported by selected providers.}{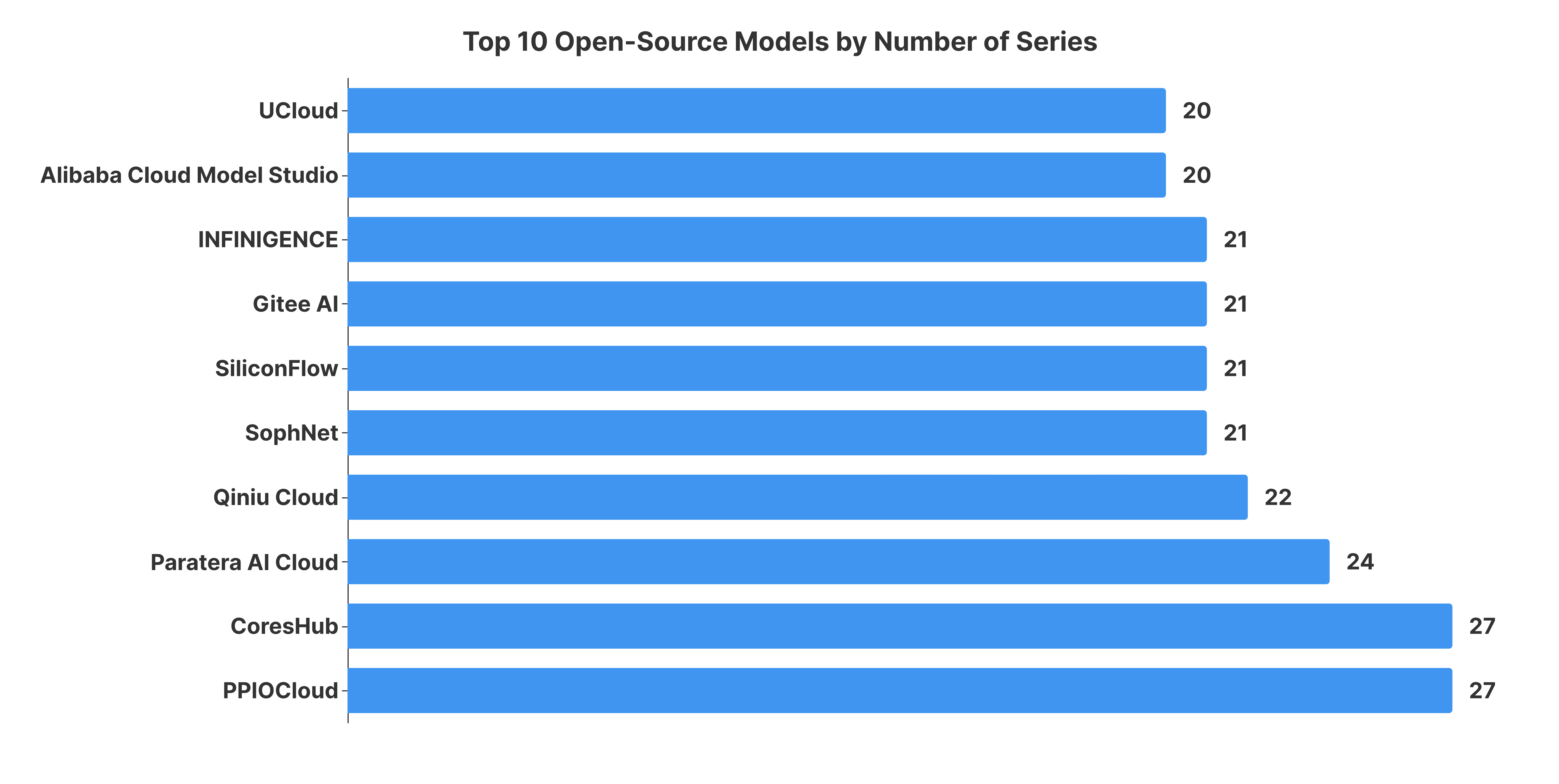}

\subsection{Anchored Prices, Dispersed Performance}
Figure~\ref{fig:7} reports relative pricing for popular models. Most entries are close to official pricing, with the chart's ``identical to official'' category defined as values within 5\% of the official benchmark. This concentration indicates that listed price is relatively anchored for popular models.

\figone{Relative pricing categories among providers for selected popular models. The categorical encoding is useful for detecting price anchoring, but numeric ratio tables are needed for independent dispersion estimates.}{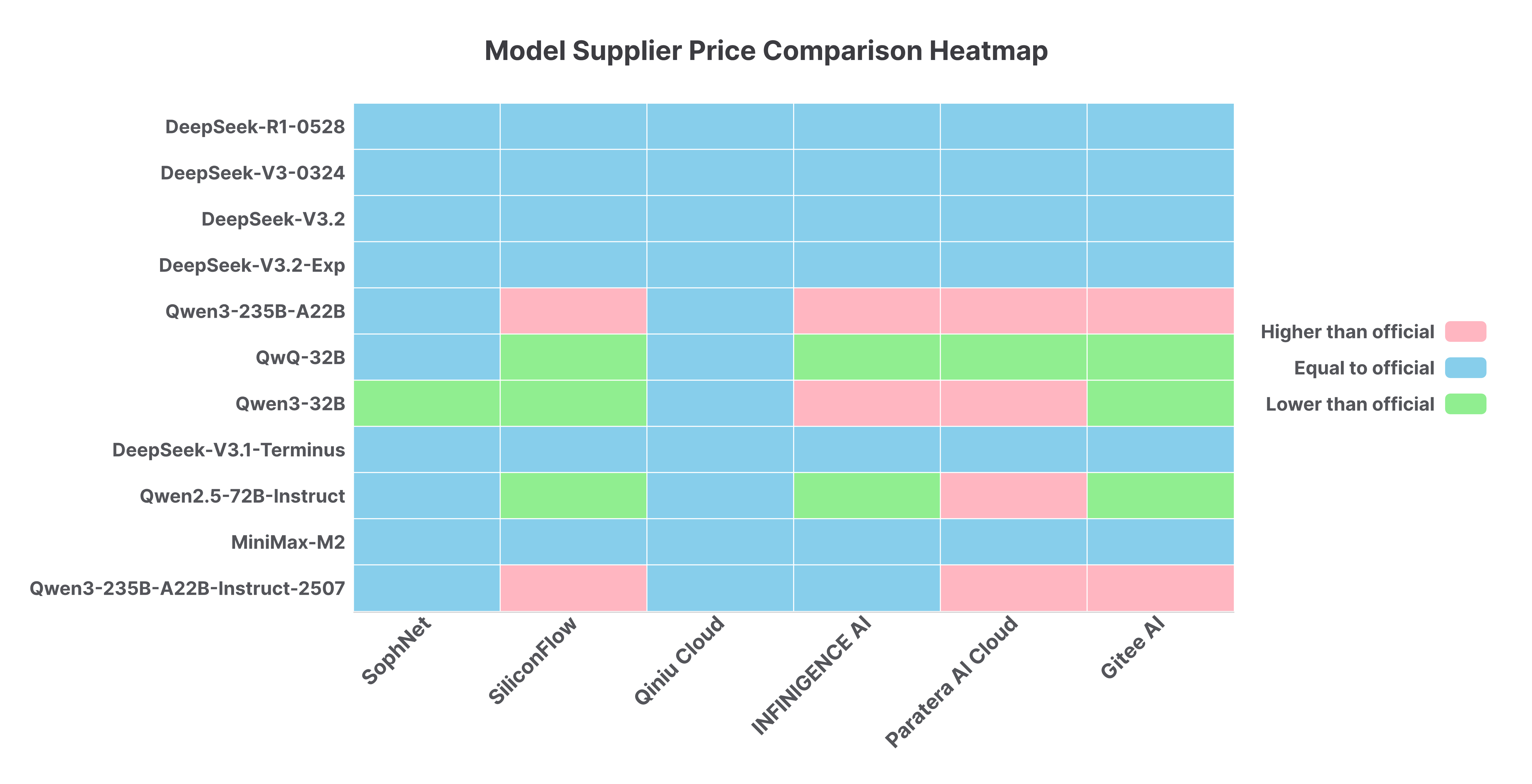}

Performance variation is substantially larger. Figure~\ref{fig:8} compares TTFT and throughput across providers and models. The same model can have different service behavior depending on provider. Official endpoints are not uniformly fastest, and third-party providers are not uniformly slower. This cross-provider spread is the empirical basis for treating service selection as a statistical decision under heterogeneity.

\figone{Provider-level performance categories in TTFT and TPS. A stronger artifact release should expose p50/p90 TTFT and TPS plus provider-model sample sizes.}{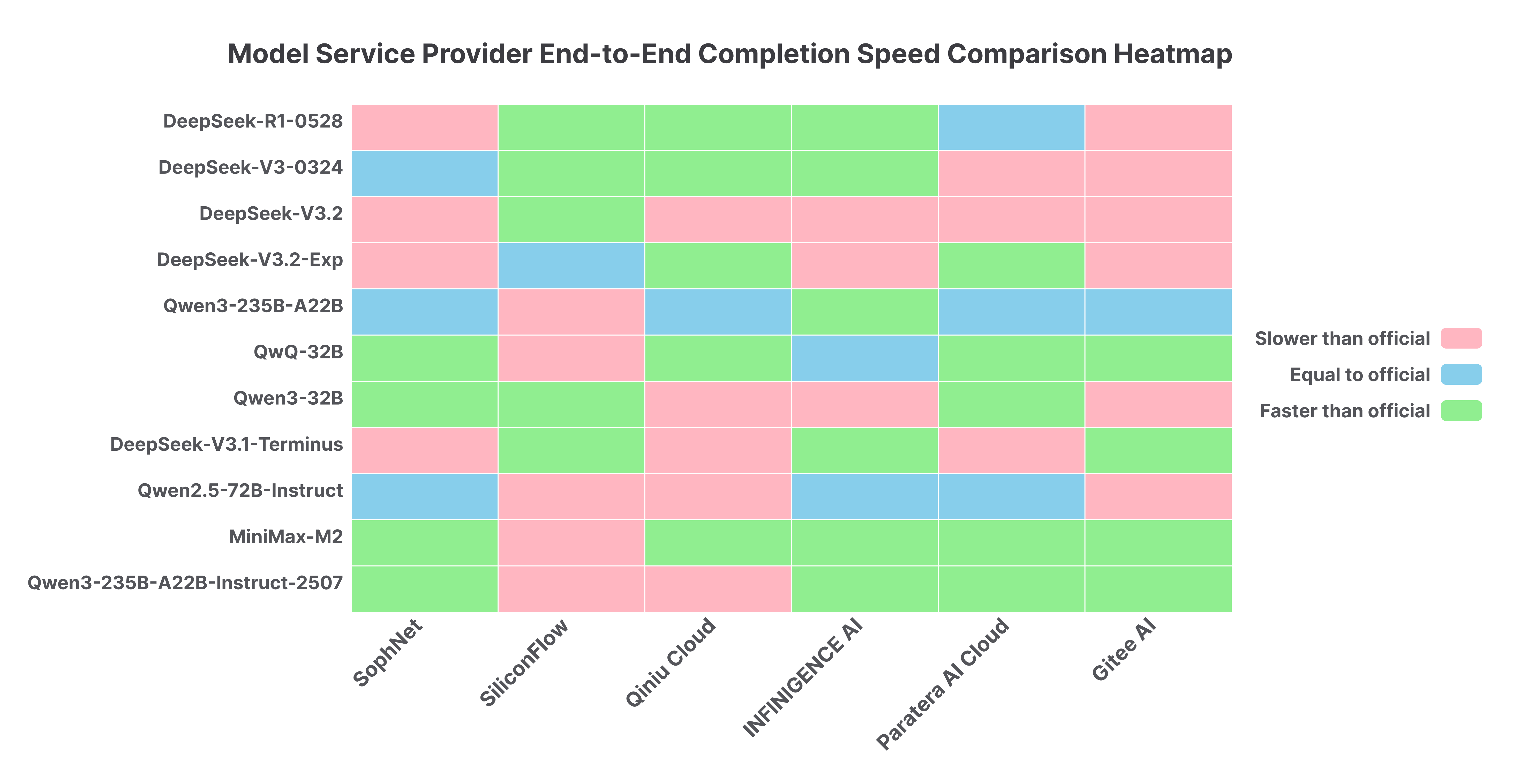}

Figure~\ref{fig:9} adds context length. Context length differs from price and speed because it has a threshold effect. For long-context retrieval, long-document question answering, compliance review, repository-level code understanding, or multi-turn customer service, a reduced context window can move an endpoint from feasible to infeasible. Thus, context shrinkage is better modeled as a constraint in the feasible set than as a continuous penalty in an objective function.

\figone{Context length variation among service providers. Absolute context-window values and tested-versus-advertised limits are required for exact reproducibility.}{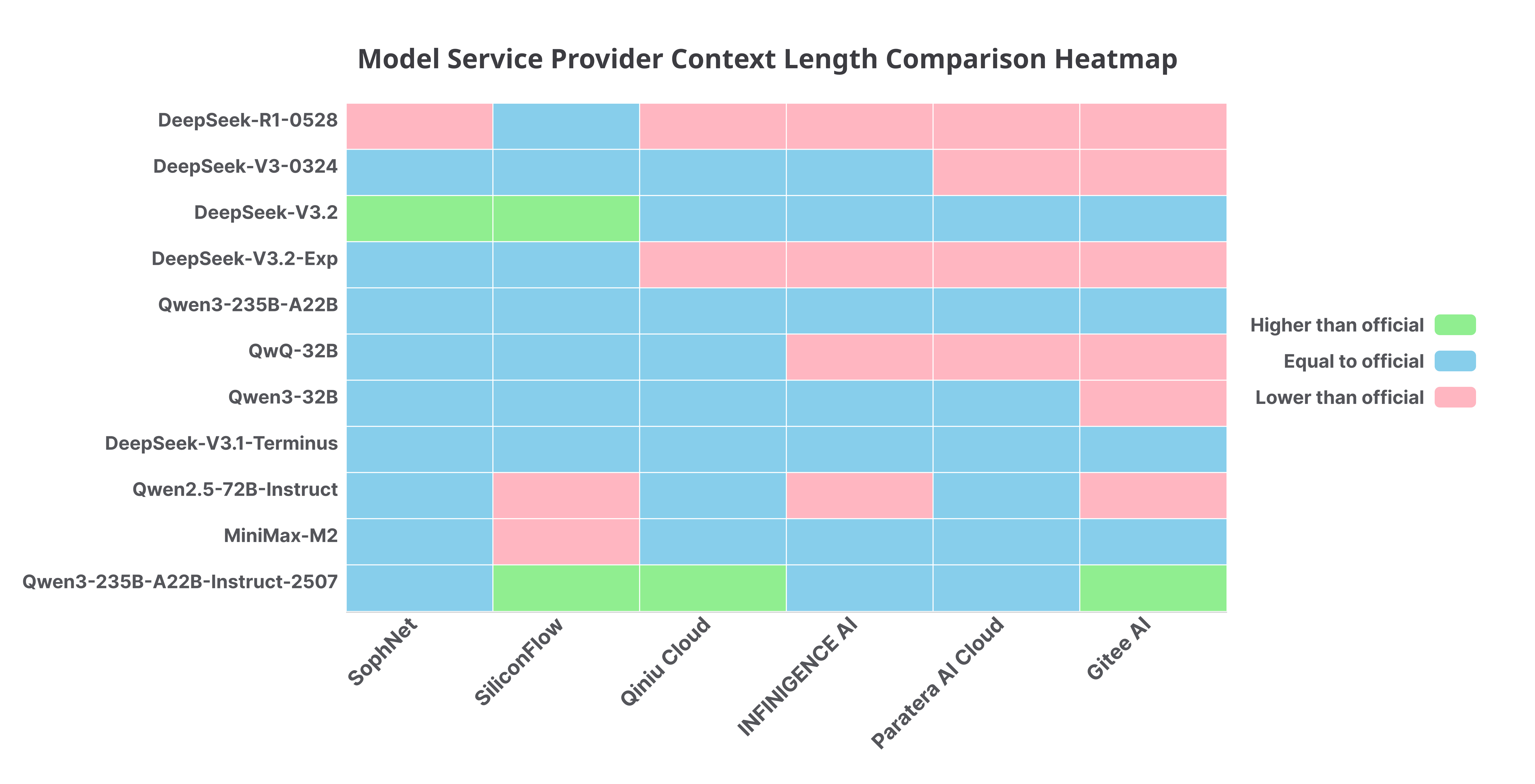}

\subsection{Service Quality Is Time-Varying}
Figure~\ref{fig:10-1} compares the first observed week after model listing with the last observed week in the collection period. Several models show lower TTFT and stable or improved TPS over time. The statistical interpretation is a service-quality drift: providers appear to optimize deployments after launch, reducing typical latency and, in some cases, narrowing the interquartile range. This means that provider performance is time-varying. A routing or provider-selection policy based on a single historical measurement can become stale.

\figone{First-week versus last-week service-quality comparison for selected provider-model observations. The figure is descriptive and should be interpreted alongside request-length and task-mix controls.}{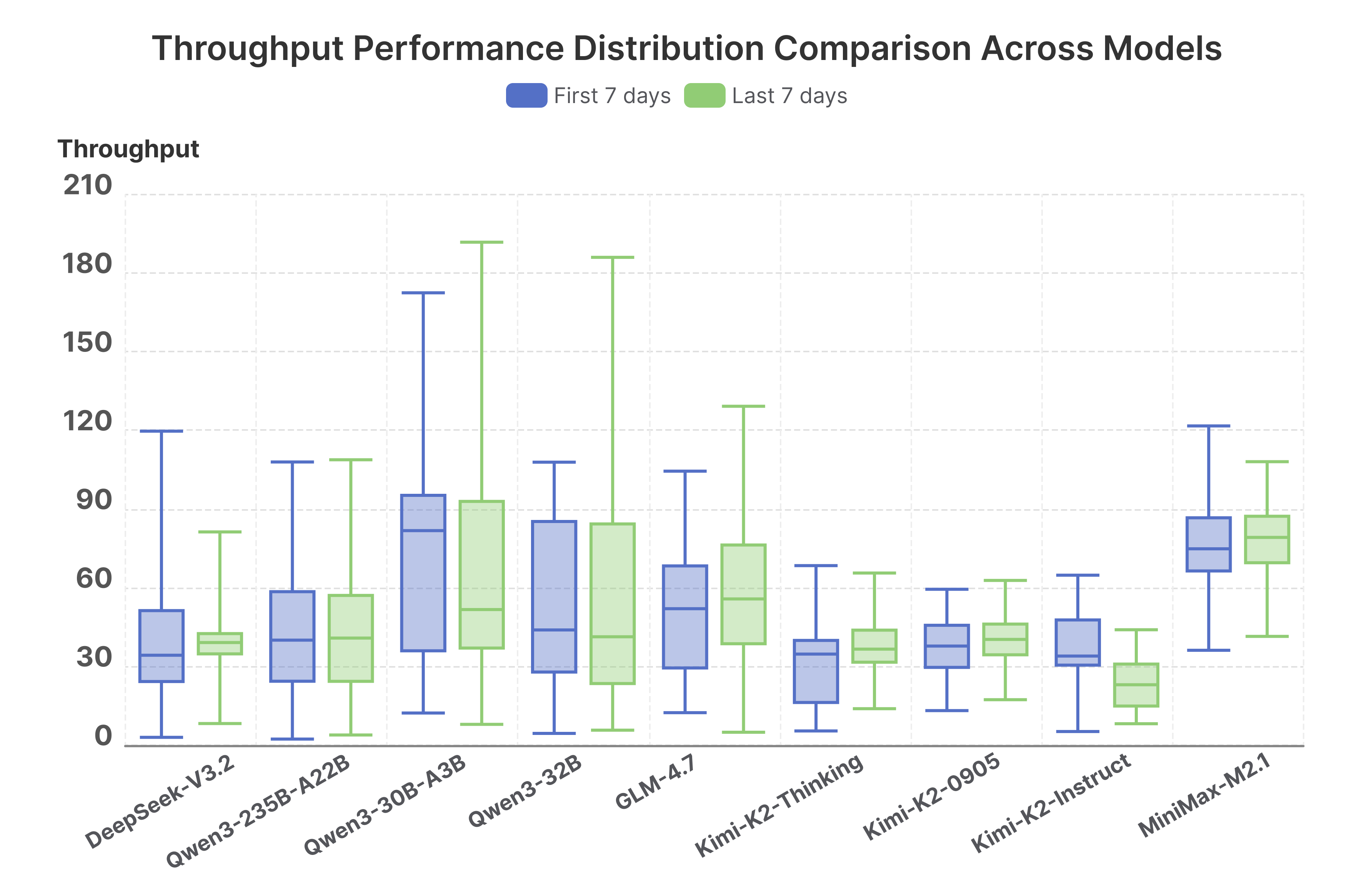}

\subsection{Protocol Compatibility as a Feasible-Set Constraint}
Figure~\ref{fig:11} evaluates compatibility with OpenAI-compatible and Anthropic-compatible interfaces across 18 providers. The OpenAI-compatible interface is much more widely supported: each tested model receives support from 12.2 providers on average. Anthropic-compatible support is lower, with an observed adoption rate of 29.7\% and an average of 3.6 providers per tested model. The highest observed Anthropic-compatible support in the tested set is 50.0\% for Kimi-K2-Thinking, while Qwen3-30B-A3B reaches 14.3\%.

\figone{Protocol support rates for OpenAI-compatible and Anthropic-compatible interfaces across the tested provider subset.}{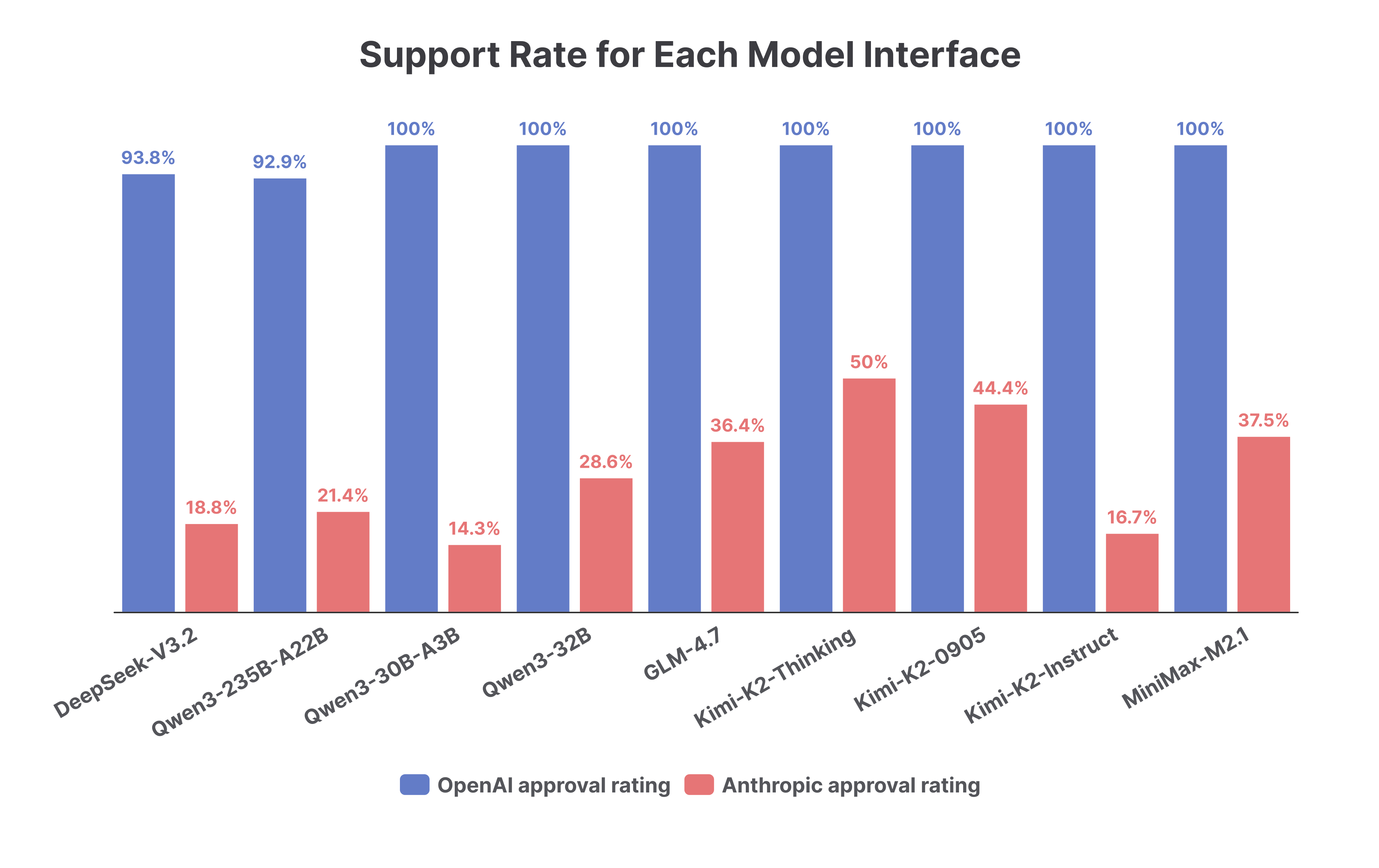}

The gap has a statistical consequence for application portability. When an application uses a less widely supported protocol, the feasible provider set shrinks. This can increase concentration risk and reduce the variance available for routing optimization. Protocol compatibility is therefore not merely an implementation detail; it changes the size and composition of the candidate set.

\subsection{Error Semantics and Recoverability}
Availability does not guarantee correctness. The observed provider APIs show inconsistent error semantics, including successful HTTP status codes with error payloads, empty bodies for failures, generic 500 errors, and streaming responses that contain error descriptions without stable fields. Table~\ref{tab:provider_api_issues} summarizes representative cases.

\begin{table}[t]
\centering
\caption{Representative provider API error-semantics issues observed in the sample.}
\label{tab:provider_api_issues}
\footnotesize
\setlength{\tabcolsep}{4pt}
\renewcommand{\arraystretch}{1.15}
\begin{tabularx}{\linewidth}{c X c X}
\toprule
\textbf{HTTP status} & \textbf{Response body} & \textbf{Observed category} & \textbf{Issue} \\
\midrule
200 & Empty body & Very high & Failure is not distinguishable from an empty successful response. \\
200 & JSON body containing an error object & Very high & Error is encoded inside a success status. \\
400 & Service-side API exception & Relatively low & Server failure is represented as a client error. \\
500 & Unknown error & Very high & The error class is too coarse for automated attribution. \\
503 & Empty body & Occasional & No stable diagnostic field is available. \\
\bottomrule
\end{tabularx}
\end{table}

These patterns increase the engineering cost of automated recovery. A robust client must infer failures from several inconsistent signals rather than from a stable status-code and schema contract. The public table is qualitative because the denominator is not released; a measurement-complete version should report the provider-model count, request count, and share of each error class. In distributional terms, errors are not independent scalar events; their observability and recoverability vary by provider.

\subsection{Tail Latency Concentrates by Provider-Model Pair}
Figure~\ref{fig:12} shows slow-response ratios by provider under the operational threshold used by the measurement system. Some providers maintain slow-response rates below 0.3\% while serving more than one million calls, indicating stable capacity planning under large load. Other providers exhibit slow-response rates approaching 5\% despite lower total volume, which is consistent with insufficient elasticity, cold starts, long scheduling chains, or poor peak-load isolation. Because the threshold is not public, the result is best used as an internal reliability category; threshold-free tail statistics are needed for independent verification.

\figone{Slow-response ratio by provider.}{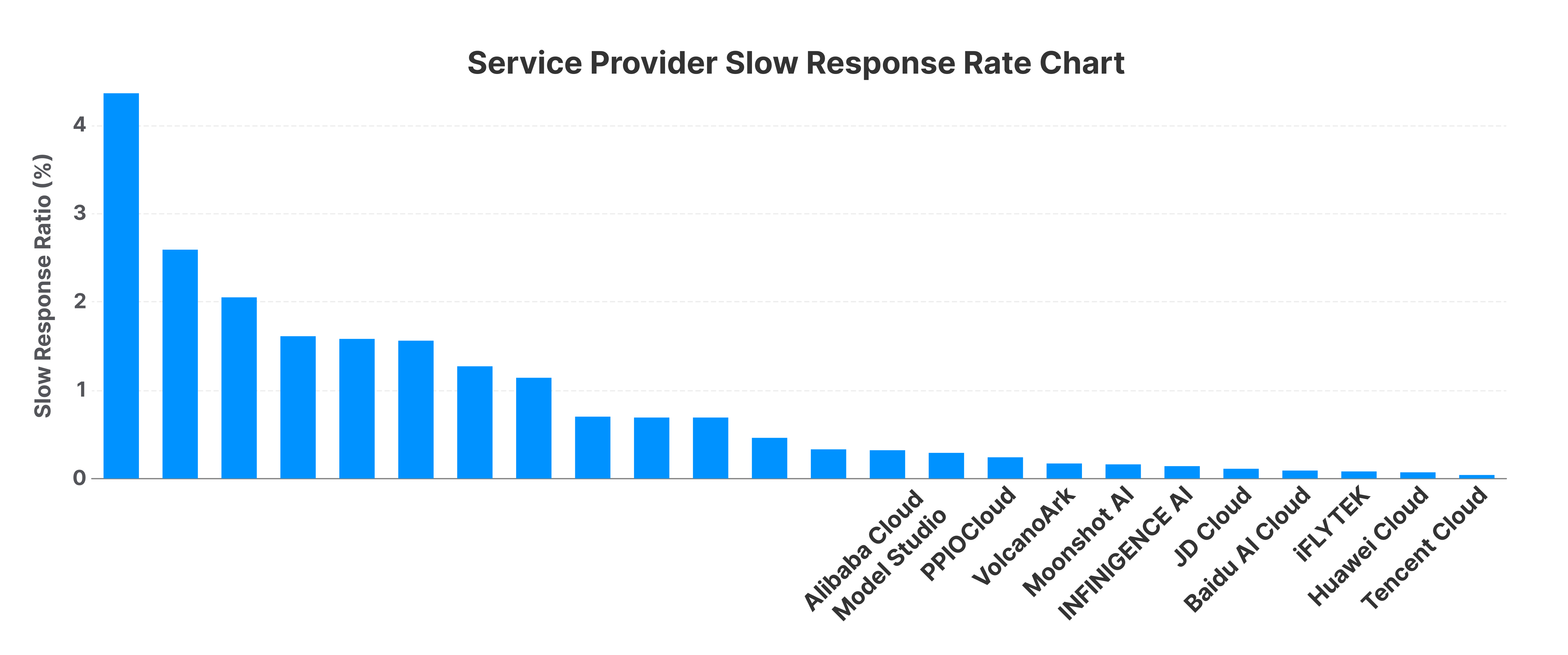}

At the model level, Figure~\ref{fig:13} shows that slow-response rates also vary by model. DeepSeek-V3.2, for example, has an observed slow-response rate of approximately 1.49\%. Because latency depends on both model inference characteristics and provider infrastructure, the appropriate unit of quality analysis is the provider-model pair, not the provider alone or the model alone.

\figone{Slow-response ratio by model.}{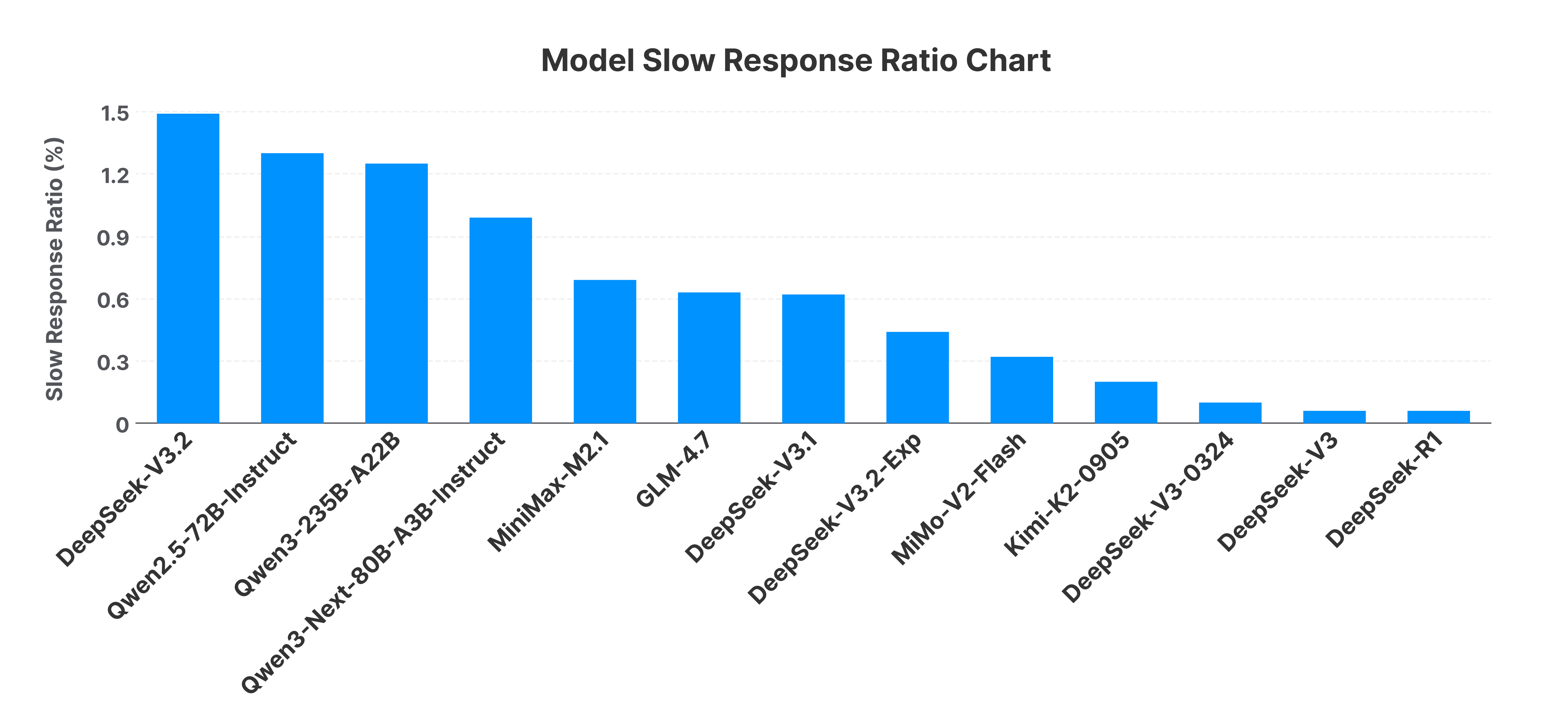}

\subsection{Breadth Versus Depth in Provider Strategy}
The provider-level evidence suggests a breadth-depth trade-off. Some providers list many models quickly, capturing broad coverage and early availability. Others allocate resources to fewer models and can achieve stronger throughput or latency on those specific models. The trade-off is statistical: broad coverage increases the probability that an arbitrary requested model is available, while deep optimization reduces variance and tail risk for selected high-volume models. A single provider count cannot capture this two-dimensional structure.

\section{Task-Conditioned Service Selection}
\subsection{Token-Length Regimes}
Figure~\ref{fig:14} plots task categories by input and output length. The categories occupy different regions rather than a single common cluster. News and information tasks lie in a long-input, short-output region, indicating retrieval-heavy or summarization-heavy workloads. Creative writing and service-assistance tasks lie closer to long-input, long-output regimes, where total token cost dominates. Content-generation tasks tend toward output-heavy generation. Professional services, knowledge translation, education and entertainment, and technology development cluster around medium or shorter inputs and outputs, where TTFT and reliability may dominate user experience.

\figone{Input and output length distribution across labeled application categories. The task taxonomy applies only to the labeled subset of traffic.}{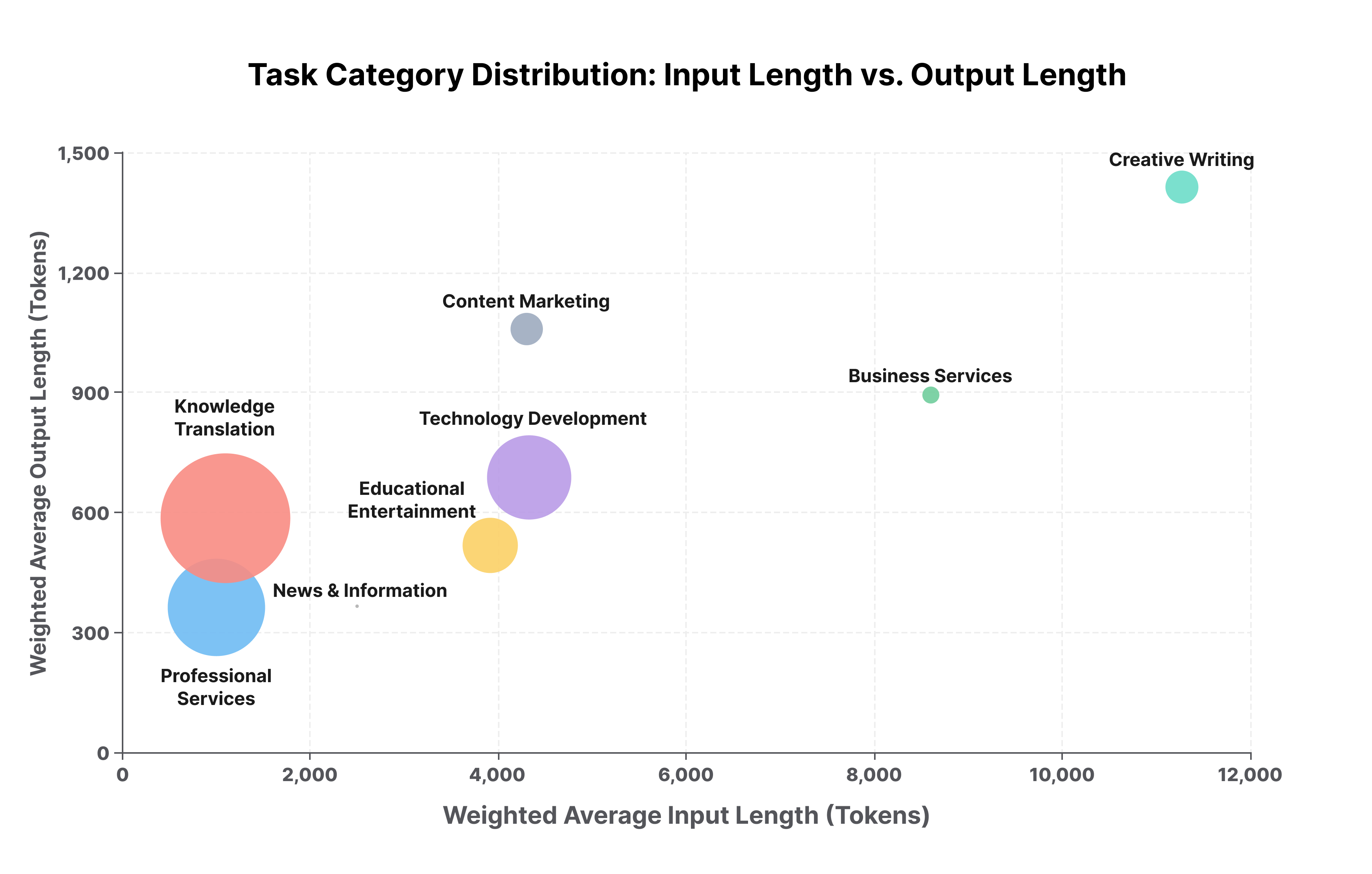}

This heterogeneity changes the correct optimization target. A long-input, short-output request is sensitive to input price, prefill speed, and context length. A short-input, long-output request is sensitive to output price and decode throughput. A short-input, short-output request may be dominated by fixed overhead and queuing latency. A long-input, long-output request requires joint cost and latency constraints. Therefore, provider selection should be conditioned on task statistics, not only on model name.

\subsection{Conditional Workload Composition}
Figure~\ref{fig:15} shows the task composition of each model family in the labeled subset. The heatmap is row-normalized by model family, so it estimates \(P(\mathrm{task}\mid\mathrm{model})\), not \(P(\mathrm{model}\mid\mathrm{task})\). It can therefore support statements about the workload regimes served by each model family, but it should not by itself be used to claim that a task prefers a particular model. The row distributions are concentrated rather than uniform: DeepSeek traffic is dominated by knowledge-translation and technology-development tasks, Qwen traffic by professional services, GLM traffic by technology-development and education/entertainment, MiniMax traffic by creative-writing tasks, and Kimi traffic by content-marketing tasks.

\figone{Row-normalized task composition by model family, estimating \(P(\mathrm{task}\mid\mathrm{model})\) within the labeled subset. A column-normalized companion table is required to estimate \(P(\mathrm{model}\mid\mathrm{task})\).}{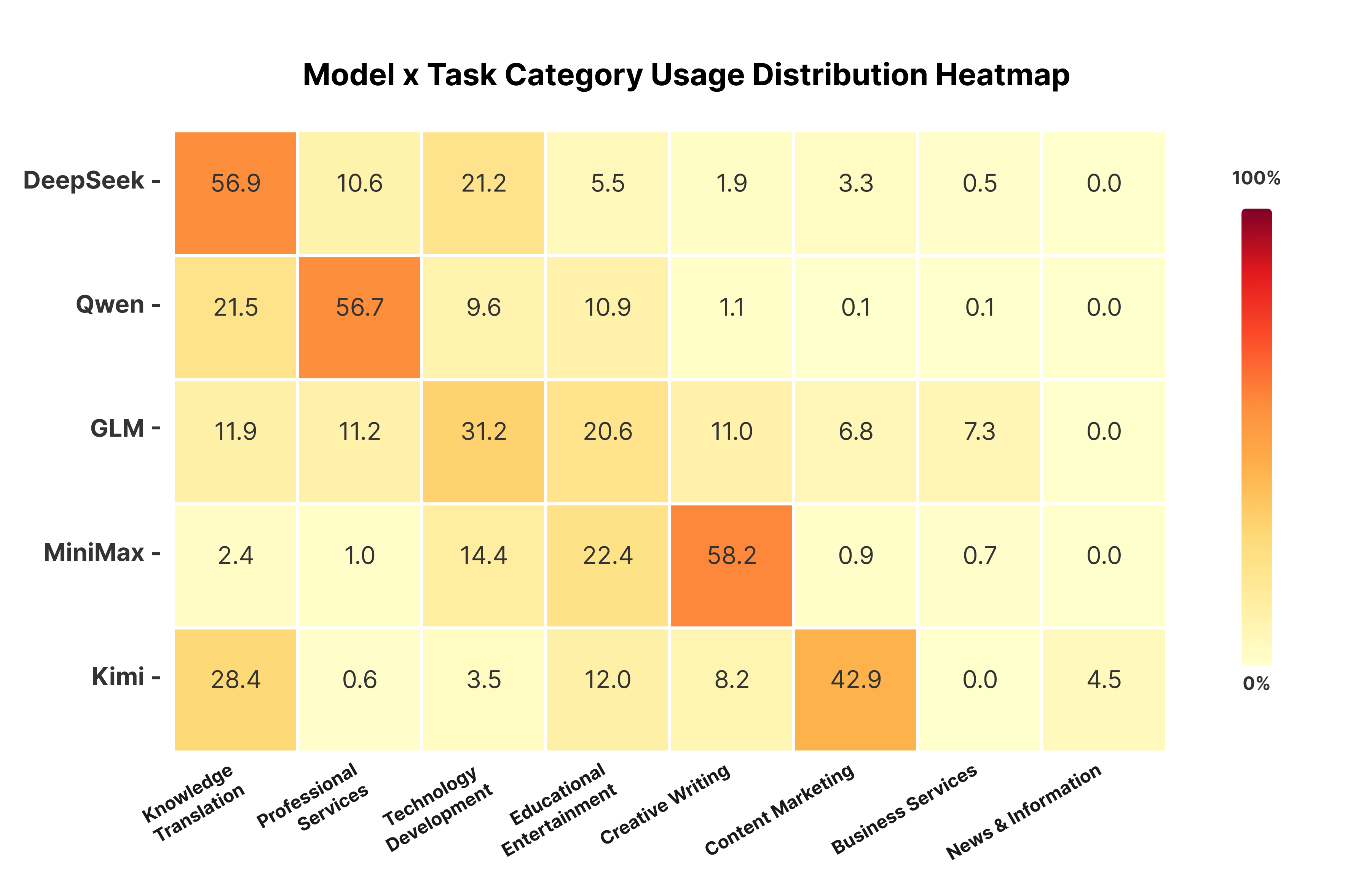}

The statistical implication is narrower but still important: aggregate model usage can hide the workload mix carried by each family. For routing, benchmarking, or provider evaluation, aggregate traffic should be decomposed by task type, token-length regime, context requirement, and normalization direction. A stronger task-specialization claim requires a companion column-normalized matrix, task-level denominators, and uncertainty estimates.

\subsection{Observed Routing Preferences}
Figure~\ref{fig:16-1} reports selected routing policies. In the observed sample, 77.1\% of default routing configurations use performance-driven strategies. The figure should be interpreted as a configuration-level distribution rather than a mutually exclusive request-level distribution unless otherwise specified by the released denominator. It is a revealed-preference signal: when users can choose among routing objectives, a large fraction prioritize speed or throughput over pure cost minimization.

\figone{Distribution of selected routing policies.}{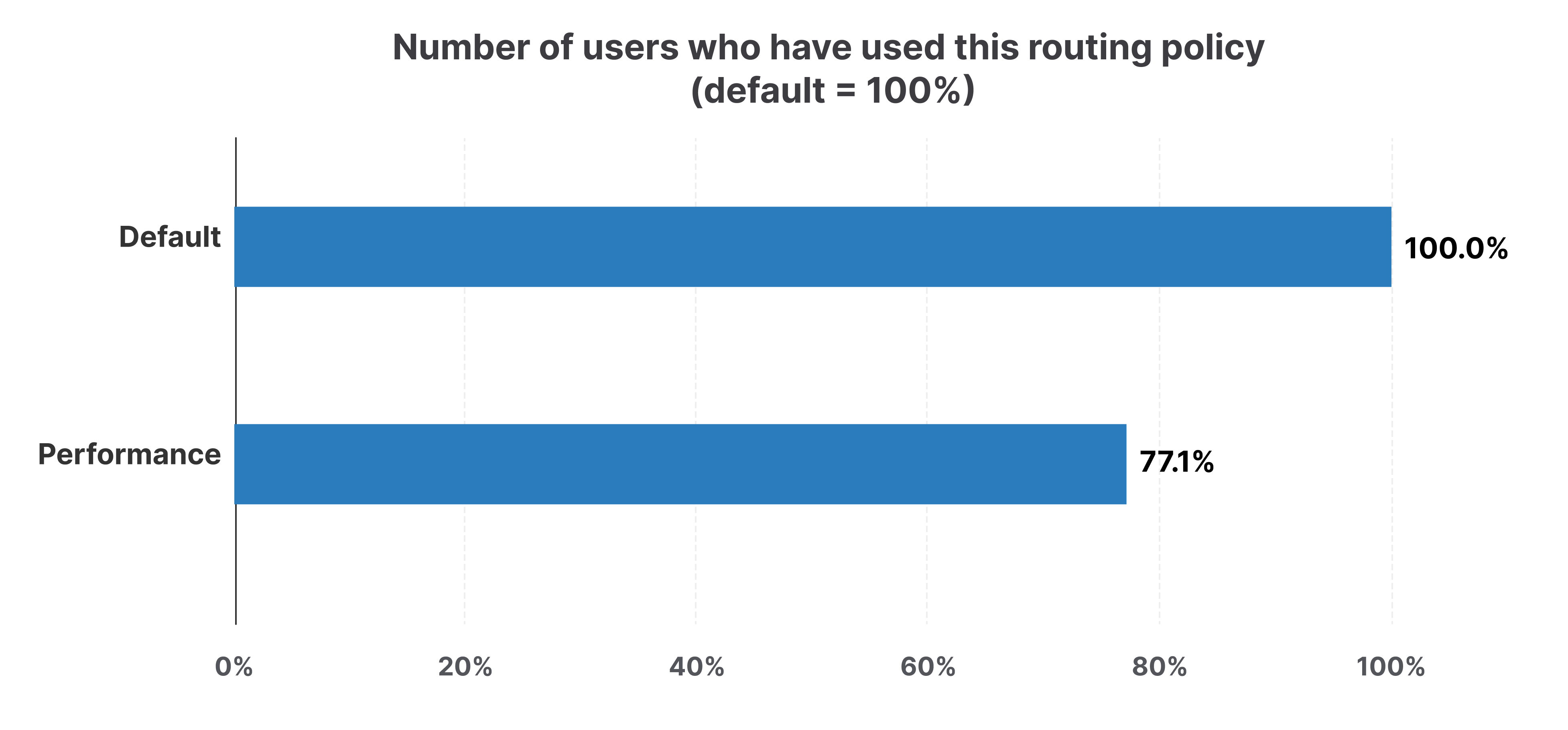}

The result should not be read as a universal preference statement for all LLM users. It is conditional on the measured population and the available provider set. Still, it aligns with the provider-level evidence: when prices are anchored but performance variance is large, the marginal value of selecting a faster provider can exceed the marginal value of selecting a slightly cheaper one.

\subsection{Counterfactual Routing Effects}
We evaluate routing using two representative counterfactual comparisons under fixed model labels. The estimand is the aggregate cost or throughput that would be observed if the same visible request population were served under a specified baseline policy and the observed feasibility constraints. This is weaker than a randomized A/B test: it does not prove semantic quality equivalence across providers, nor does it eliminate all queue-state or time-varying confounding. It is nevertheless useful for estimating the operational value of measured provider dispersion.

For Qwen3-32B, official pricing is used as the baseline and routed provider selection is evaluated on approximately 1.5 million requests, more than 1 billion input tokens, and about 660 million output tokens. The observed routed cost is 4,577 yuan, compared with 7,355 yuan under official pricing, corresponding to a 37.8\% reduction. This estimate should be read as a price-schedule counterfactual over the observed token mix. A complete routing evaluation should additionally report retry cost, failure rate, truncation rate, tool-call or JSON-schema conformance, and a task-specific quality proxy.

\figone{Counterfactual cost comparison for Qwen3-32B.}{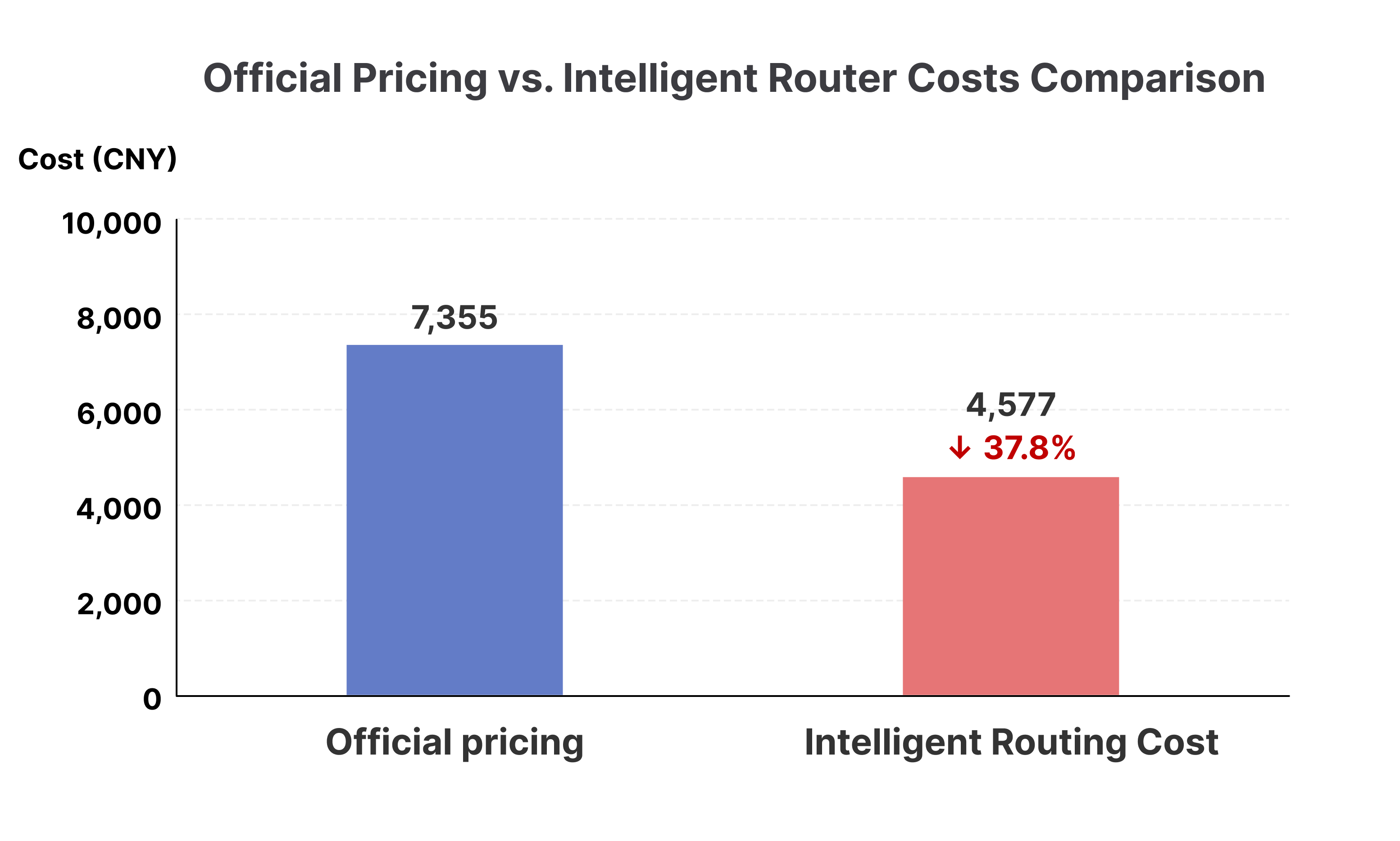}

For DeepSeek-V3.2, throughput is compared between official access and routed access on approximately one million requests. The observed average TPS under routing is about 90\% higher than the official baseline, with especially large gains for requests generating more than 1,000 tokens. Figure~\ref{fig:18} shows the distributional shift: routing reduces low-throughput cases and increases the mass of high-throughput cases. The figure should be interpreted as throughput evidence, not full service equivalence evidence; p50/p90/p99 TPS, TTFT, error rates, and output-length-stratified shares are required for a complete measurement artifact.

\figone{Throughput comparison between official access and routed access for DeepSeek-V3.2.}{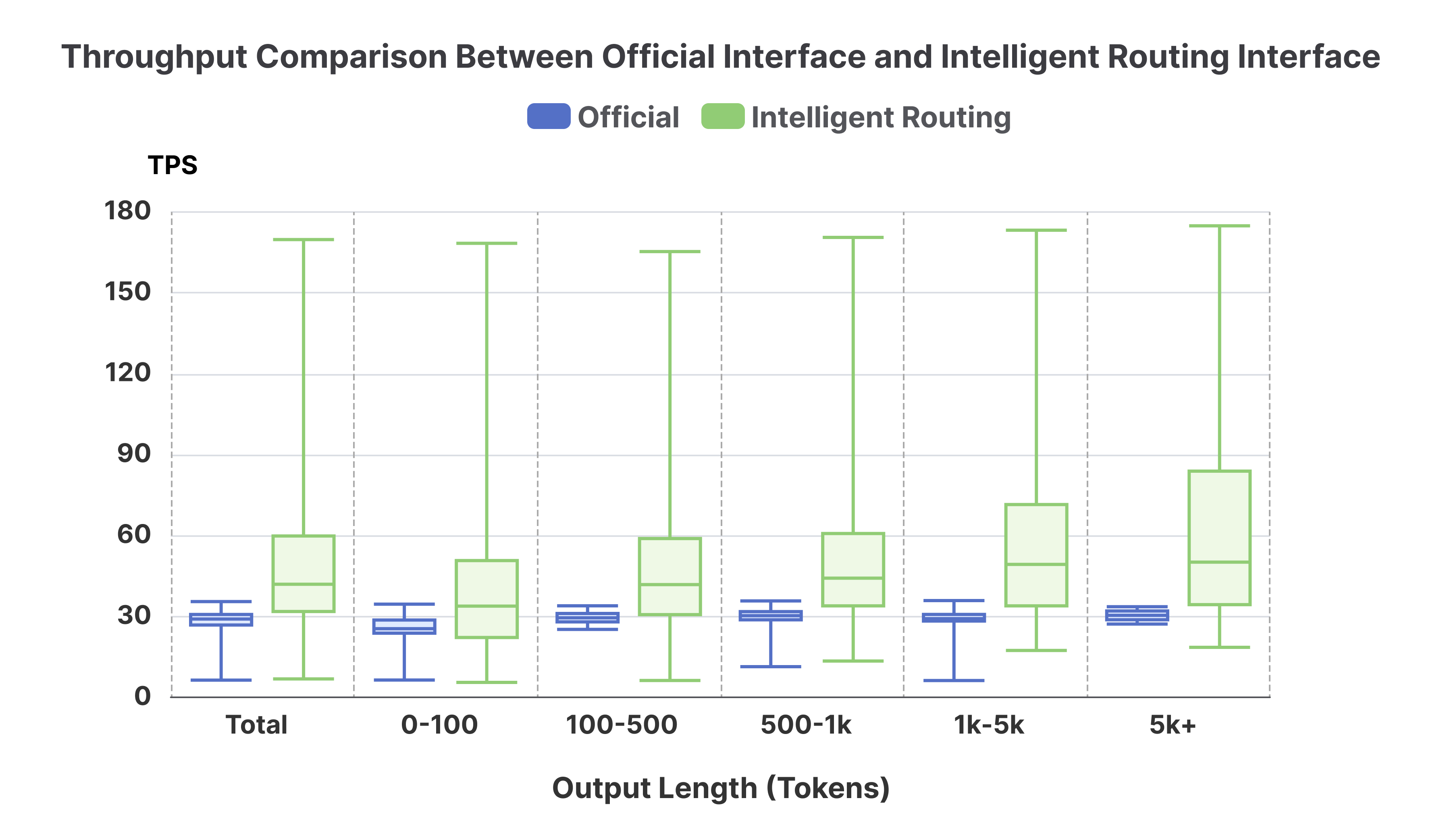}

These comparisons are not claims that routing always improves every request. They show that when provider performance is heterogeneous and measurable, a routing policy can improve aggregate outcomes by selecting from the upper tail of the provider distribution while respecting price and availability constraints. The appropriate baselines for a stronger follow-up are direct official access, cheapest feasible provider, fastest feasible provider, random feasible provider, static best feasible provider, task-conditioned routing, and an oracle upper bound computed on held-out windows.

\subsection{Temporal and Geographic Load Structure}
Figure~\ref{fig:19} shows a daily cycle in request volume and active users. Activity declines from late night to early morning, reaches a trough around 6--8 AM, rises during work hours, peaks around 3--4 PM, and forms a secondary evening plateau. User activity also rebounds around 9--11 PM without a proportional increase in request volume, suggesting lower per-user call intensity during late evening interactive use.

\figone{Daily request and user activity, normalized to noon.}{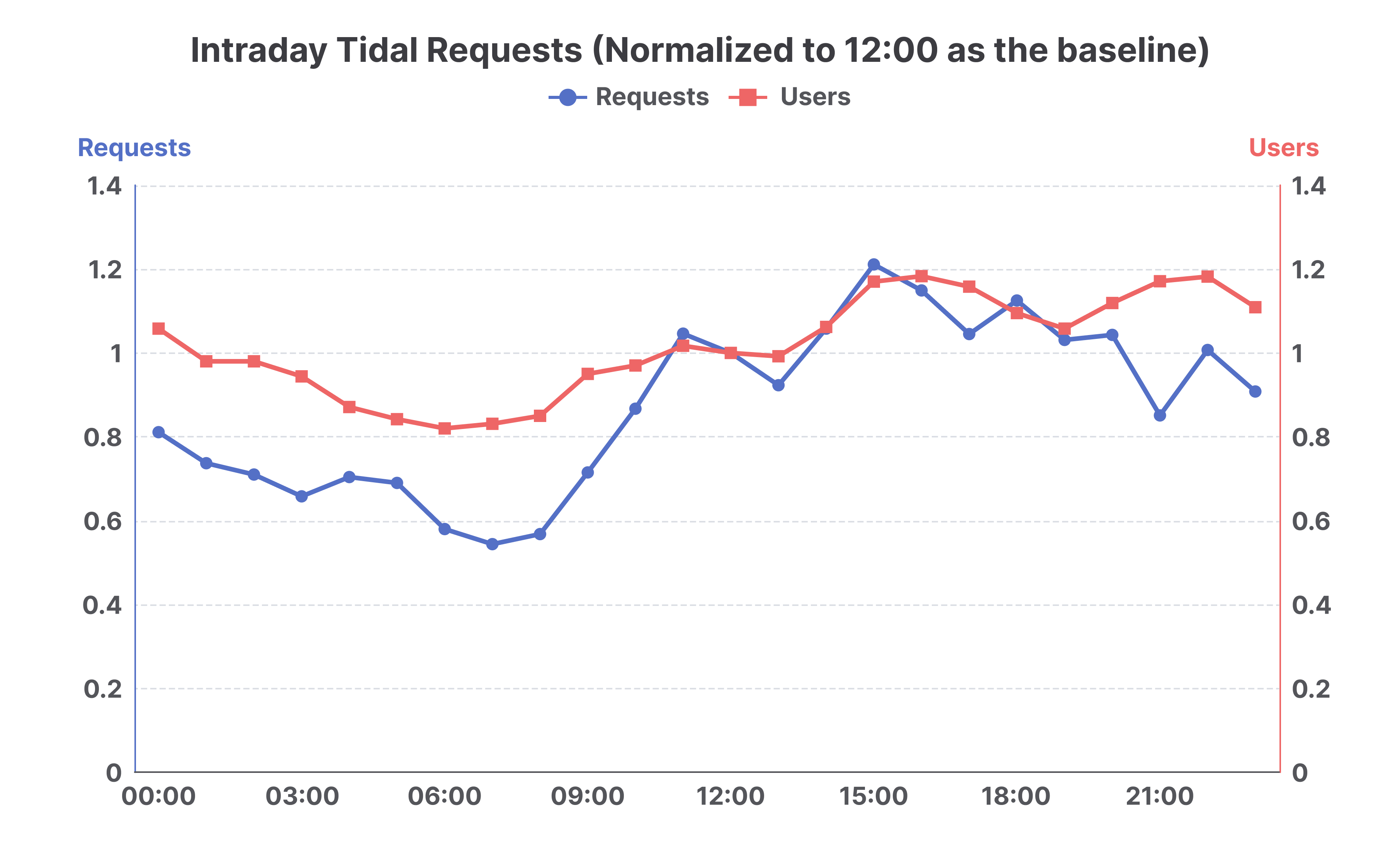}

Figure~\ref{fig:20-1} shows the weekly distribution. Weekdays contain stronger localized hotspots, while weekends are more diffuse. This pattern suggests that batch or workflow-driven calls contribute substantially to weekday peaks, whereas weekend usage is less concentrated. The normalization is relative to the displayed temporal aggregate and should not be interpreted as an absolute traffic forecast.

\figone{Weekly request distribution, normalized to noon on Monday.}{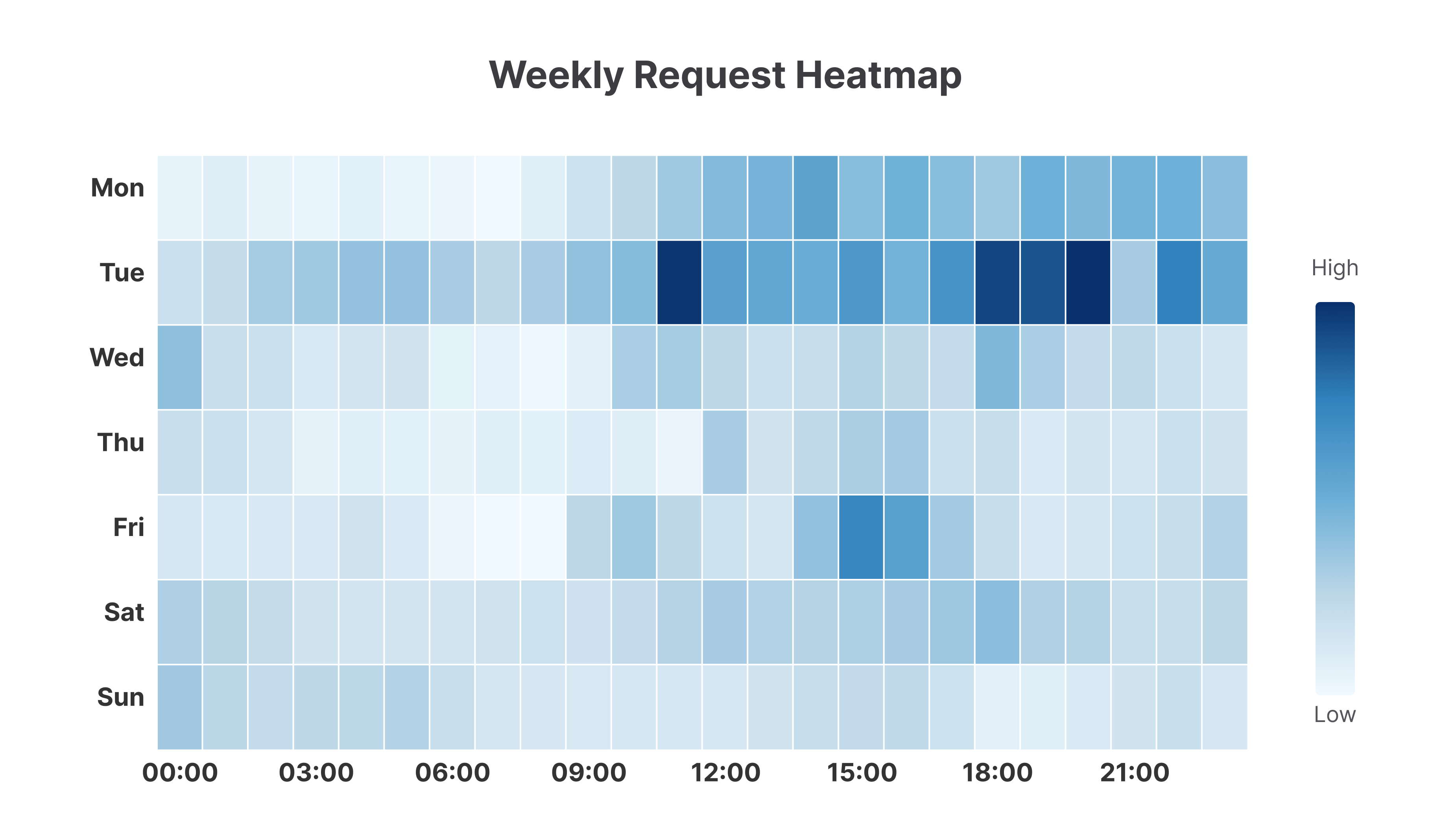}

Figure~\ref{fig:21-1} summarizes geographic distribution by time period. In the observed sample, domestic requests dominate, with Beijing and other domestic regions contributing 46.3\% and 42.8\%, respectively; overseas requests account for 10.9\%. The Beijing series has a stronger afternoon-to-evening peak, which may reflect concentrated batch jobs or workflow triggers from a smaller number of high-volume users. Because the geographic variable can reflect request ingress, account region, network vantage point, or provider routing rather than end-user location, the figure is treated as workload context rather than a market-geography estimate.

\figone{Geographic distribution of requests by time period.}{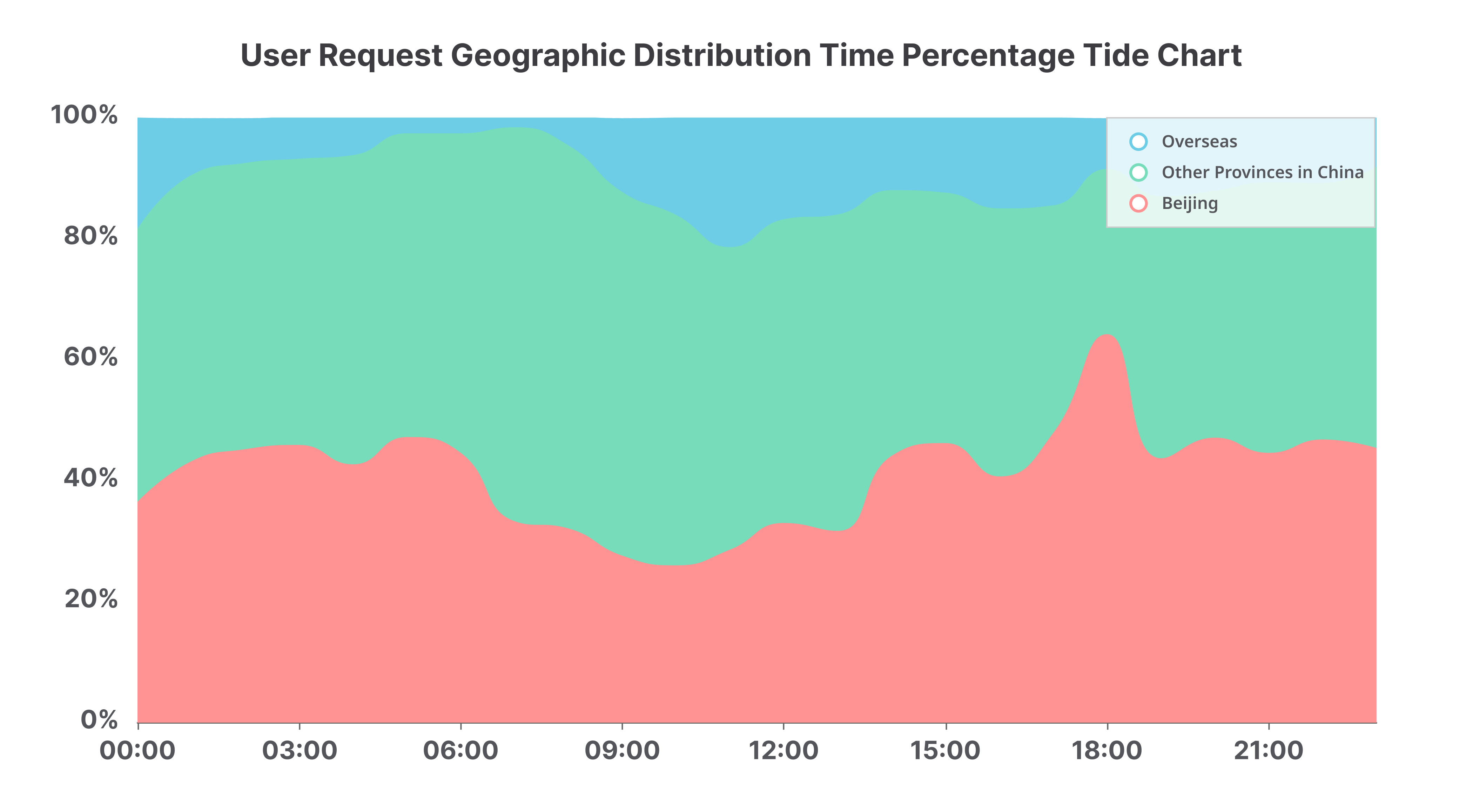}

\section{Discussion}
The measurements point to a simple but consequential lesson: in hosted LLM inference, a model name is not a sufficient description of the service being purchased or operated. The same model family can have different availability, context capacity, protocol behavior, latency, and throughput depending on the provider and the time of measurement. Treating ``DeepSeek-V3.2'' or ``Qwen3-32B'' as a single operational object hides the variation that applications actually experience.

Demand concentration and version persistence also need to be understood together. The head of the model distribution is narrow, but it is not a pure winner-take-all race toward the newest release. Older versions continue to receive traffic when they have already been validated inside applications. For practitioners, this means that migration pressure and benchmark improvements do not translate automatically into production replacement. For providers, it means that removing or degrading an older endpoint can affect more real traffic than release chronology alone would suggest.

Provider coverage is similarly easy to misread. Listing a model creates an option, but it does not prove demand; high traffic proves adoption, but it does not prove broad supply. The interesting cases are mismatches: widely listed models with little traffic, and high-traffic models whose performance depends strongly on provider choice. These mismatches are where operational risk and routing value appear.

Price is the most visible attribute of an API endpoint, but it is not the dimension with the largest observed operational spread. In the high-volume part of the sample, listed prices often remain near reference levels, while TTFT, TPS, context windows, protocol compatibility, and error semantics vary enough to change whether an endpoint is usable for a task. Context length is especially important because it is not a mild quality adjustment: below a task's required window, the endpoint leaves the feasible set.

The routing results should therefore be read as a measurement of exploitable dispersion. Routing is useful not because one endpoint is always best, but because the feasible provider set contains time-varying tails: lower latency, higher throughput, lower cost, or more reliable protocol behavior can be selected when those quantities are continuously measured. This also explains why task conditioning matters. A short interactive request, a long-output generation request, and a long-context retrieval request expose different bottlenecks; aggregating them into one average hides the decision boundary that a real router needs.

\section{Data Governance}
\subsection{Privacy and Anonymization}
The request-level corpus comes from real API operations and is therefore treated as sensitive operational data. The paper uses aggregated statistics, normalized counts, categorical task labels, and provider-model-level measurements. It does not publish prompts, completions, account identifiers, IP addresses, API keys, or customer-specific time series. Geographic and temporal results are reported only as aggregate distributions.

\subsection{Artifact Availability}
The public release accompanying this manuscript contains the \LaTeX{} source, aggregate figures, and a minimal set of derived artifacts that do not expose prompts, completions, API keys, or account-level traces. The intended artifact boundary is: aggregate CSV tables for public figure values; provider-metadata schema; pricing snapshot table; compatibility probe templates; latency-probe schema; routing-counterfactual summary; and figure-reproduction scripts where raw inputs are public. For figures derived from private traffic logs, absolute counts are normalized or omitted, but the normalization direction, denominator type, and aggregation level should be stated. A synthetic replay trace can be released to exercise the routing simulator without exposing customer traffic.

\subsection{Conflict of Interest}
Several authors are affiliated with QingCheng.AI AIPing Team, and the measurement data were collected through the \aiping{} service environment. This creates a potential institutional conflict of interest. The routing results are therefore limited to the measured counterfactual samples and should not be read as a claim that any provider or policy is uniformly optimal.

\section{Threats to Validity}
The analysis is based on \dataset{}, not a complete population census. The measured user base, provider set, and task composition may differ from the broader ecosystem. Some prices were manually collected at historical points and may not reflect discounts, promotions, or private contracts. Task categories are available only for a subset of requests. Compatibility tests cover selected protocols and models rather than every possible endpoint. The counterfactual routing comparisons fix model labels but cannot remove all differences in request mix, queue state, provider-side temporal variation, or unobserved quality drift.

Model identity is a separate threat. A provider-visible name does not prove checkpoint equality. Providers may quantize, fine-tune, watermark, add safety layers, transform prompts, change decoding defaults, or route internally to related variants. This paper therefore avoids checkpoint-level equality claims unless the measurement design can support them. Behavioral fingerprinting or formal model-equality tests are needed to move from nominal identity to distributional identity.

Aggregation is another threat. Family-level groupings can hide variant-level behavior, especially when instruction, thinking, dated, and experimental variants are grouped together. The main text uses family aggregation for public summary figures, but variant-level sensitivity analyses are required before using those aggregates for fine-grained operational claims. Finally, figures that report normalized request volume preserve relative structure but do not disclose absolute totals, limiting independent re-estimation of statistics that require the undisclosed full population. Concentration indices computed from public normalized bars should therefore be interpreted as statistics over the displayed model-family set.

\section{Conclusion}
This paper presents a measurement study of open-weight LLM API services from Q4 2025. The main conclusion is that the hosted open-weight ecosystem should be analyzed as a distribution of service objects rather than a catalog of model names. Model demand is concentrated but persistent across versions; provider coverage is broad but unevenly aligned with realized traffic; listed prices are comparatively anchored while latency, throughput, context length, compatibility, and error semantics vary materially. Application categories introduce additional heterogeneity through token-length regimes and workload composition. For application builders, the cheapest or newest endpoint is rarely a sufficient decision rule. For service operators and researchers, the measured object is the provider-model-task-time tuple, and robust evaluation requires sample-frame disclosure, identity boundaries, feasibility constraints, and uncertainty around headline statistics.

\appendix
\section{Model-Family Aggregation Rules}
\label{app:classification}
\begin{itemize}
  \item \textbf{DeepSeek series models}
    \begin{itemize}
      \item \textbf{DeepSeek-V3/R1}: DeepSeek-V3; DeepSeek-R1; DeepSeek-V3-0324; DeepSeek-R1-0528.
      \item \textbf{DeepSeek-V3.1}: DeepSeek-V3.1; DeepSeek-V3.1-Terminus.
      \item \textbf{DeepSeek-V3.2}: DeepSeek-V3.2; DeepSeek-V3.2-Speciale; DeepSeek-V3.2-Exp.
    \end{itemize}
  \item \textbf{Qwen series models}
    \begin{itemize}
      \item \textbf{Qwen3-235B-A22B}: Qwen3-235B-A22B; Qwen3-235B-A22B-Instruct-2507; Qwen3-235B-A22B-Thinking-2507.
      \item \textbf{Qwen3-Coder-480B-A35B}.
      \item \textbf{Qwen3-32B}.
      \item \textbf{Qwen3-30B-A3B}: Qwen3-30B-A3B; Qwen3-30B-A3B-Instruct-2507; Qwen3-30B-A3B-Thinking-2507.
      \item \textbf{Qwen3-VL-235B}: Qwen3-VL-235B-A22B-Instruct; Qwen3-VL-235B-A22B-Thinking.
      \item \textbf{Qwen3-VL-30B}: Qwen3-VL-30B-Instruct; Qwen3-VL-30B-Thinking.
      \item \textbf{Qwen2.5-72B}.
      \item \textbf{QwQ-32B}.
    \end{itemize}
  \item \textbf{Kimi series models}
    \begin{itemize}
      \item \textbf{Kimi-K2}: Kimi-K2-Thinking; Kimi-K2-0905; Kimi-K2-Instruct.
    \end{itemize}
  \item \textbf{GLM series models}
    \begin{itemize}
      \item \textbf{GLM-4.6}.
      \item \textbf{GLM-4.7}.
    \end{itemize}
  \item \textbf{MiniMax series models}
    \begin{itemize}
      \item \textbf{MiniMax-M2}.
      \item \textbf{MiniMax-M2.1}.
    \end{itemize}
\end{itemize}

\section{Terminology}
\begin{description}[style=nextline,leftmargin=2.6cm]
  \item[TPS] Tokens per second, measured as the average output-token rate during generation.
  \item[TTFT] Time to first token, measured as the time from request submission to the first streamed token.
  \item[Throughput] Output tokens completed per unit time in a measurement window.
  \item[Slow response] A response whose latency exceeds the operational threshold used in the measurement system.
  \item[HTTP status semantics] A 4xx status conventionally indicates a client-side error, while a 5xx status indicates a server-side error.
\end{description}

\section{Recommended Identity and Sensitivity Checks}
\label{app:identity}
The title of this paper intentionally stresses the gap between model name and service behavior. A complete checkpoint-level identity audit requires additional tests beyond the public aggregate figures. The recommended protocol is:
\begin{enumerate}
  \item Fix a prompt set, decoding parameters, streaming mode, tokenizer, request schema, and client region.
  \item Collect repeated samples from each provider endpoint and from a reference implementation when the checkpoint is available.
  \item Compare distributions using a two-sample test such as the MMD-style model-equality tests proposed in prior work~\cite{modelequality}.
  \item Report separate results for checkpoint-level, provider-visible variant-level, and family-level aggregation.
  \item Repeat core demand, coverage, and routing statistics under no aggregation, variant-level aggregation, and family-level aggregation.
\end{enumerate}
These checks are not required for the descriptive provider-performance claims in the main text, but they are required before making strong claims that two providers are serving the same checkpoint distribution.

\section{Reproducibility Checklist}
\label{app:repro}
For a stronger public release, each figure should be backed by a machine-readable artifact with the following columns where applicable: measurement window, aggregation level, provider identifier or anonymized provider code, model variant, model family, protocol, context limit, input tokens, output tokens, completion status, retry status, TTFT, TPS, end-to-end latency, price schedule, task label, normalization denominator, and bootstrap unit. Routing-counterfactual artifacts should additionally include baseline policy, routed policy, feasibility filters, request bucket, token bucket, observed cost, baseline cost, success rate, truncation rate, and any quality proxy used.

\end{document}